\documentclass[aps,pra,letterpaper,superscriptaddress,twocolumn,showpacs,floatfix,10pt]{revtex4}
\usepackage{amsfonts}
\usepackage{amsmath}
\usepackage{amssymb}
\usepackage{graphicx}
\usepackage{epstopdf}
\usepackage{epsfig}
\usepackage{color}
\newcommand{\be}{\begin{equation}}
\newcommand{\ee}{\end{equation}}
\newcommand{\bea}{\begin{eqnarray}}
\newcommand{\eea}{\end{eqnarray}}
\newcommand{\ba}{\begin{array}}
\newcommand{\ea}{\end{array}}

\newcommand{\eq}[1]{(\ref{#1})}

\newcommand{\Tr}{\mbox{Tr}}

\newcommand{\al}{\alpha}
\newcommand{\bt}{\beta}

\newcommand{\om}{\omega}

\newcommand{\cH}{{\cal{H}}}

\newcommand{\ket}[1]{|{#1}\rangle}
\newcommand{\ga}{\gamma}

\begin{document}

%%%%%%%%%%%%%%%%%%%%%%%%%%

\title{Symmetry Protected Quantum State Renormalization}

\author{Ching-Yu Huang}
\affiliation{Max-Planck-Institut f$\ddot{u}$r Physik komplexer
Systeme, 01187 Dresden, Germany}
\email{ayajoa@pks.mpg.de}

\author{Xie Chen}
\affiliation{Department of Physics, University of California at Berkeley, Berkeley, CA
94720, USA}
\email{xiechen@berkeley.edu}

\author{Feng-Li Lin}
\affiliation{Department of Physics, National Taiwan Normal University, Taipei, 116, Taiwan}
\email{linfengli@phy.ntnu.edu.tw}

\vfill
\begin{abstract}

Symmetry protected topological (SPT) phases with gapless edge excitations have been shown to exist in strongly interacting bosonic/fermionic systems and it is highly desirable to identify practical systems realizing such phases through numerical simulation. A central question to be addressed is how to determine the SPT order in the system given the simulation result while no local order parameter can be measured to distinguish the phases from a trivial one. In the tensor network approach to simulate strongly interacting systems, the quantum state renormalization algorithm has been shown to be effective in identifying intrinsic topological orders. Here we show that a modified algorithm can identify SPT orders by extracting the symmetry protected fixed point entanglement pattern in the ground state wave function which is essential for the existense of SPT order. The key to this approach is to add proper symmetry protection to the renormalization process. We demonstrate the effectiveness of this algorithm with examples of nontrivial SPT phases with internal symmetry in 1D and internal and translation symmetry in 2D.
\end{abstract}

\maketitle

\tableofcontents

%%%%%%%%%%%%%%%%%%%%%%%%%%%%%%%%%%%%%%%%%%%%%%%%%%%%%%%%%%%%%%%%%%%%%%%%%%%%%%%%%%%%%%
\*\\
 \section{Introduction}

% SPT order, exactly solvable models in strongly interacting systems

Symmetry protected topological (SPT) phases are bulk-gapped quantum
phases with symmetries \cite{spt}. If the system is on a closed
manifold, the ground state does not spontaneously break the
symmetry. On the other hand, if the system has a boundary, there are
gapless or degenerate edge states as long as the symmetries are not
explicitly broken. Therefore SPT phases represent a nontrivial type
of order beyond Landau's symmetry breaking theory. Topological
insulators and superconductors are examples of SPT phases in free
fermion systems
\cite{Hasan2010,Qi2011,Hasan2011,Schnyder2008,free1,free2}. In one
spatial dimension, a complete understanding of all possible SPT
phases in interacting systems has been
obtained\cite{1daklt1,1daklt2,1dBartlet2010,1dspt, Schuch2011,
Turner2011, Fidkowski2011, Pollmann2012} starting from the classic
example of Haldane phase in spin 1 chains\cite{Haldane1983,
AKLT1987}. Recently, it has been discovered that nontrivial SPT
orders can also exist in strongly interacting boson/fermion systems
in two and higher dimensions \cite{spt,fSPT}. Exactly solvable
models were presented which has a gapped and symmetric bulk and
gapless symmetry protected edge states
\cite{2dexact1,2dexact2,fSPT}.

% Where to find practical systems to realize such phases? Numerical simulation of strongly interacting systems. Tensor network approach.

It is highly desirable to find such strongly interacting SPT phases
in experiments, similar to their free fermion counterparts. While
the exactly solvable models prove the existence of SPT orders in
strongly interacting systems, it is very unlikely that such models
can be realized in experiments as they usually involve multi-body (6 or 7-body) interactions. In order to determine which
physically realistic systems can have SPT order, numerical
simulations are necessary. The tensor network renormalization
algorithm \cite{Verstraete2008, Verstraete2009, Vidal2010, trg,
Gu2008,Singh2012} is a powerful and generic approach to simulate
strongly interacting boson/fermion systems in two and higher
dimensions and therefore can play a major role in the discovery of
strongly interacting SPT orders.

% How to identify topological order, symmetry protected in particular, from the numerical simulation data? No local order parameter to measure.

A major question to be addressed in the tensor network approach to
simulate SPT phases is how to identify the SPT order. Symmetry
breaking phases can be identified by measuring local order
parameters in the ground states. However, as SPT ground states do
not break any symmetry, no local measurement can distinguish an SPT
phase from a trivial symmetric phase. Systems with intrinsic
topological order (like $Z_2$ spin liquids or fractional quantum
Hall systems) on the other hand can be identified by measuring the
ground state degeneracy on a torus \cite{deg1,deg2} or the
topological entanglement entropy \cite{tee1,tee2}. However, these
quantities are both trivial for SPT states.

% Ground state entanglement is the key. long/short range entanglement. quantum state RG is effective in identifying intrinsic topological order/long range entanglement. However, straightforward application removes short range entanglement. We modify the algorithm by adding symmetry protection when doing RG and demonstrate that the short range entanglement structure of certain SPT state can be properly protected.

An important signature of SPT phases is the existence of nontrivial entanglement structure in their ground states \cite{spt,2dexact1}. Compared to trivial symmetric phases whose ground states can be simple product states (for example the $\prod(|\uparrow\rangle+| \downarrow\rangle)$ state in the transverse field Ising model), the entanglement structure in the ground states of SPT phases cannot be totally removed as long as symmetry is not broken. Therefore, SPT ground states are characterized by short-range entanglement which is protected by symmetry. This is similar to systems with intrinsic topological orders where the long-range entanglement patterns in the ground states are essential for the existence of the order. It has been shown that the long-range entanglement patterns can be effectively extracted using a quantum state renormalization algorithm based on the tensor network representation of the ground states \cite{2drg,jj}. Can similar ideas be applied to SPT orders?

% In particular, we demonstrate that the algorithm works for 1D and 2D AKLT

Naively, one might expect that the quantum state renormalizaton algorithm fails to distinguish SPT order from trivial symmetric phases, as the algotirhm is designed to remove short range entanglement structures from the state and retain only the long-range one. After removing all short range entanglement, the ground state of SPT phases becomes a total product state which is the same as a trivial symmetry state. However, this is only possible if the symmetry of the system is broken during the process. If we require that symmetry is always preserved during the renormalization procedure, some short-range entanglement structures are always kept which can be used to identify the SPT order at the renormalization fixed point.

% The paper is organized as follows

In this paper, we show how such a symmetry protection can be properly added to the quantum state renormalizaiton algorithm. The algorithm we propose applies to general 1-dimensional (1D) SPT phases protected by internal symmetry and 2-dimensional (2D) weak SPT phases protected by both internal and translation symmetry. As an example, we demonstrate the effectiveness of our algorithm by applying it to the 1D and 2D AKLT phases and show that the SPT order in these systems can be successfully identified from the fixed point short-range entanglement pattern of the states.

The paper is organized as follows: in section \ref{review}, we review the notion of SPT order (especially that in AKLT states) and the quantum state renormalization algorithm which can be used to identify intrinsic topological orders; section \ref{SPQSRG} discusses how symmetry protection can be added to the quantum state renormalization procedure and how it can be used to identify the SPT order in 1D and 2D AKLT states which allows us to determine the phase diagram of 1D and 2D anti-ferromagnetic spin models; in section \ref{conclusion} we conclude our discussion and talk about open questions. In Appendix \ref{app a} we briefly review the notion of projective representation; some numerical results of solving 2D AKLT-like model are given in Appendix \ref{app b}, and the explicit form of its fixed point tensor in Appendix \ref{app c}. We want to emphasize that, even though we use as examples the AKLT states with $SO(3)$ symmetry, our algorithm works equally well for SPT orders with any other symmetry group.

%%%%%%%%%%%%%%%%%%%%%%%%%%%%%%%%%%%%%%%%%%%%%%%%%%%%%%%%%%%%%%%%%%%%%%%%%%%%%%%%%%%%%%%%%%%%%%%%%%%%%%%%%%%%%%%%%%%%%%%%%%%%%%%%%%%%%%%%%%%%%%%%
%%%%%%%%%%%%%%%%%%%%%%%%%%%%%%%%%%%%%%%%%%%%%%%%%%%%%%%%%%%%%%%%%%%%%%%%%%%%%%%%%%%%%%%%%%%%%%%%%%%%%%%%%%%%%%%%%%%%%%%%%%%%%%%%%%%%%%%%%%%%%%%%

\section{Review}
\label{review}

\subsection{Symmetry protected topological order}
\label{SPT}

% summary SPT order

Symmetry protected topological (SPT) order is characterized by the robust gapless edge modes of a bulk gapped phase without intrinsic topological order or spontaneously symmetry breaking. These gapless edge modes are protected as long as the symmetry of the system is unbroken.

% in one dimension, due to projective rep. example: spin 1 Heisenberg chain, 1D AKLT

\begin{figure}[htbp]
\begin{center}
\includegraphics[scale=0.4]{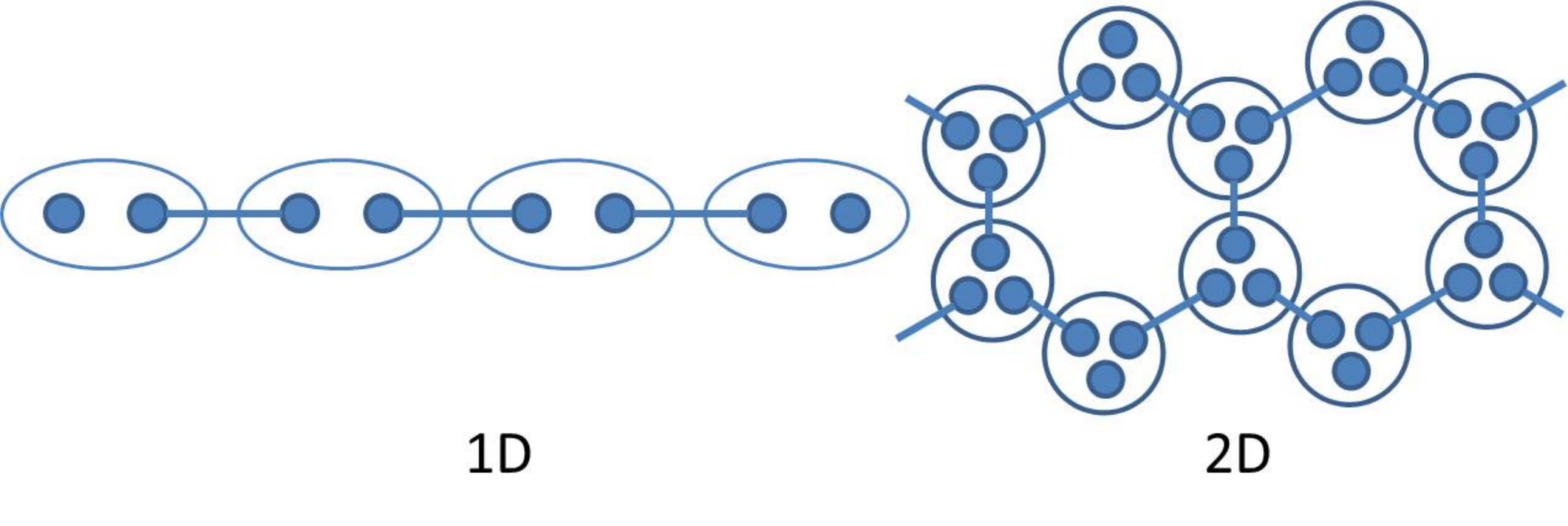}
\end{center}
%Fig 1
\caption{Valence bond structure of 1D SPT states and 2D weak SPT states. Each dot carries a projective representation of the internal symmetry and connected dots represent entanlged singlet pairs. In 1D,  symmetry acts linearly on each site (big circle) while in 2D on-site symmetry representation depends on the lattice.
}
\label{VBS}
\end{figure}

In one dimensional systems, such a protection has been well understood as coming from the projective representation carried by the edge degree of freedom.\cite{1daklt1,1daklt2,Schuch2011,Turner2011,Pollmann2012} Consider a system with global internal symmetry $G$. It was realized that, at large enough length scale (much larger than correlation length), the ground state of all 1D SPT phases has a valence bond structure as shown in Fig.\ref{VBS} (left). The degrees of freedom on each site (big circle) splits into two parts (small dots) which form entangled pairs (connected dots) with degrees of freedom on a neighboring site. Each dot carries a projective representation of $G$ while the total representation on each site is linear. Matrices $u(g)$ form a projective representation of symmetry group $G$ if
\begin{align}
 u(g_1)u(g_2)=\om(g_1,g_2)u(g_1g_2),\ \ \ \ \
g_1,g_2\in G.
\end{align}
where $\om(g_1,g_2)$ are $U(1)$ phase factors. If $\om(g_1,g_2)=1$, the representation is linear.
A key property of nontrivial projective representations is that the representation space cannot be one dimensional. See Appendix \ref{prorep} for a brief introduction to projective representation. Each entangled pair is invariant under the symmetry, therefore the bulk of the system is symmetric and gapped. However, on the boundary there are unpaired projective edge states, which are degenerate under the protection of symmetry.

The simplest example of a 1D SPT phase is the spin $1$ antiferromagnetic Heisenberg chain with $SO(3)$ spin rotation symmetry $H=\sum_i \vec{S}_i\vec{S}_{i+1}$.\cite{Haldane1983}
For this state, the degree of freedom on each site (spin $1$) transforms linearly under the symmetry. In the ground state, the spin $1$ effectively splits into two projective parts (spin $1/2$) which form singlets between neighboring sites. The key feature of this so-called `Haldane phase' is the existence of edge spin $1/2$ which leads to protected edge degeneracy.\cite{AKLT1987}

% 2D topo insulator / superconductor, edge modes protected against disorder. weaker version, 2D AKLT.

The valence bond picture of 1D SPT states can be generalized to
2D, giving rise to 2D SPT phases protected by both
internal and translation symmetry. Such SPT orders are usually said to be `weak' compared to the `strong' ones which are stable even with disorder. With $SO(3)$ and translation symmetry, for example, a representative 2D weak SPT state is the 2D AKLT state\cite{AKLT1988} defined on a
honeycomb lattice. Each lattice site contains one spin $3/2$ and the spin $3/2$ can
be decomposed into three spin $1/2$'s each forming a spin singlet
with another spin $1/2$ on a neighboring site, as shown in Fig.
\ref{VBS} (right). The bulk of the system is again symmetric and gapped. On the edge of
the system, there is a chain of spin $1/2$'s left uncoupled.
Perturbations to the edge may couple these spin $1/2$'s. However, as
long as the perturbation preserves $SO(3)$ and translation
symmetry, the edge state is gapless
\cite{LSM}. Therefore, the 2D AKLT state has SPT order protected by $SO(3)$ and translation symmetry.

The SPT order in the 1D and 2D AKLT states remains when the on-site Hilbert space is expanded to contain two or three spin $1/2$'s instead of just the symmetric subspace of spin $1$ and spin $3/2$. With this modification, the wave function is a product of singlets on each link and we call such states the dimer state.

\subsection{Quantum state renormalization group}
\label{sec:qsrg}

The Quantum State Renormalization Group (QSRG) transformation acts on a quantum state and
aims to extract the universal property of the system from the ground state wave function. The basic
idea is to remove non-universal short range entanglement structures
related to the microscopic details of the system from the wave
function before coarse graining. After many rounds of QSRG the
original ground state can flow to a simpler fixed-point state, from which the phase the system belongs to can be identified.

Such a QSRG algorithm was first demonstrated for 1D quantum
states based on the matrix product state representation \cite{1drg}
\begin{equation}
\label{MPS}
|\psi\rangle = \sum_{i_1,i_2,...,i_N}
\Tr (A^{i_1}A^{i_2}...A^{i_N})|i_1i_2...i_N\rangle
\end{equation}
where $i_k=1...d$ with $d$ being the physical dimension of a
spin at each site, $A^{i_k}$'s are $\chi\times \chi$
matrices related to the physical state $\ket{i_k}$ with $\chi$ being the inner dimension of
the MPS.

To implement the QSRG on the matrix product state, construct the double tensor as
\be\label{1dqsrg-i}
\mathbb{E}_{\alpha\gamma,\beta\delta}=\sum_{i} A^i_{\alpha\beta} \times (A^i_{\gamma\delta})^*
\ee
The double tensor is just the transfer matrix for calculating MPS expectation values.
Treat $\mathbb{E}$ as
a $\chi^2\times \chi^2$ matrix with row index $\alpha\gamma$ and
column index $\beta\delta$. Combine the double tensor of the two sites together into
$\tilde{\mathbb{E}}=\mathbb{E} \mathbb{E}$. Then think of
$\tilde{\mathbb{E}}_{\alpha\gamma,\beta\delta}$ as a matrix
with row index $\alpha\beta$ and column index
$\gamma\delta$. It is easy to see that with such a
recombination, $\tilde{\mathbb{E}}$ is a positive matrix and
can be diagonalized
\be
 \tilde{\mathbb{E}}_{\alpha\gamma,\beta\delta}
=\sum_{\tilde i} \lambda_{\tilde i} V_{\tilde i,\al\bt} V^*_{\tilde i,\ga\delta} ,
\ee
where we have kept only the non-zero eigenvalues $\lambda_{\tilde i}$ and the corresponding eigenvectors $V_{\tilde i,\al\bt}$.
$\tilde A$ is then given by
\be\label{1dqsrg-f}
 \tilde A^{\tilde i}_{\alpha\beta}=\sqrt{\lambda_{\tilde i}} V_{\tilde i,\al\bt},
\ee
which are the matrices representing the renormalized state.

An important property of $\mathbb{E}$ is that it uniquely determines the matrices, and hence the state, up to a local change of basis on each site \cite{NieC00,PVW0701}. That is, if $\mathbb{E}_{\alpha\gamma,\beta\delta}=\sum_{i}
A^i_{\alpha\beta} \times (A^i_{\gamma\delta})^*=\sum_{i} B^j_{\alpha\beta} \times (B^j_{\gamma\delta})^*$,
then $A^i_{\alpha\beta}$ and $B^j_{\alpha\beta}$ are related by a unitary transformation $U$:
$B^j_{\alpha\beta}= \sum_i U_{ji} A^i_{\alpha\beta}$. Therefore, from the above procedure we know that the renormalized state can be obtained from the original state with local unitaries on every two sites
\begin{align}
|\psi\rangle \rightarrow  |\tilde{\psi}\rangle= U_{1,2}\otimes
U_{3,4} \otimes ... \otimes  U_{2i-1,2i} \otimes ... |\psi\rangle.
\end{align}
By setting the range of $\tilde i$ to be over only the nonzero $\lambda$'s, $U$ removes local entanglement in an optimal way. This is
similar to the disentangler in the multi-scale entanglement
renormalization ansatz (MERA) \cite{mera} before the
coarse-graining. After several rounds of QSRG transformation, all gapped MPS flows to a fixed point form and the procedure can be performed without any truncation to $\chi$.

An analogous QSRG procedure exists for 2D quantum states based on the tensor product representation\cite{2drg}
\begin{equation}
|\psi\rangle=\sum_{i_1,i_2,...i_m...}\text{tTr}(T^{i_1}T^{i_2}...T^{i_m}...)|i_1
i_2 ... i_m...\rangle \label{TPS}
\end{equation}
where $T^i_{\alpha\beta\gamma...}$ is a local tensor with physical index $i$ and internal
indices $\alpha\beta\gamma$ etc. $\text{tTr}$ denotes tensor contraction of all the connected
inner indices according to the underlying lattice structure. Without loss of generality, we consider the honeycomb lattice here.

\begin{figure}[ht]
\center{\epsfig{figure=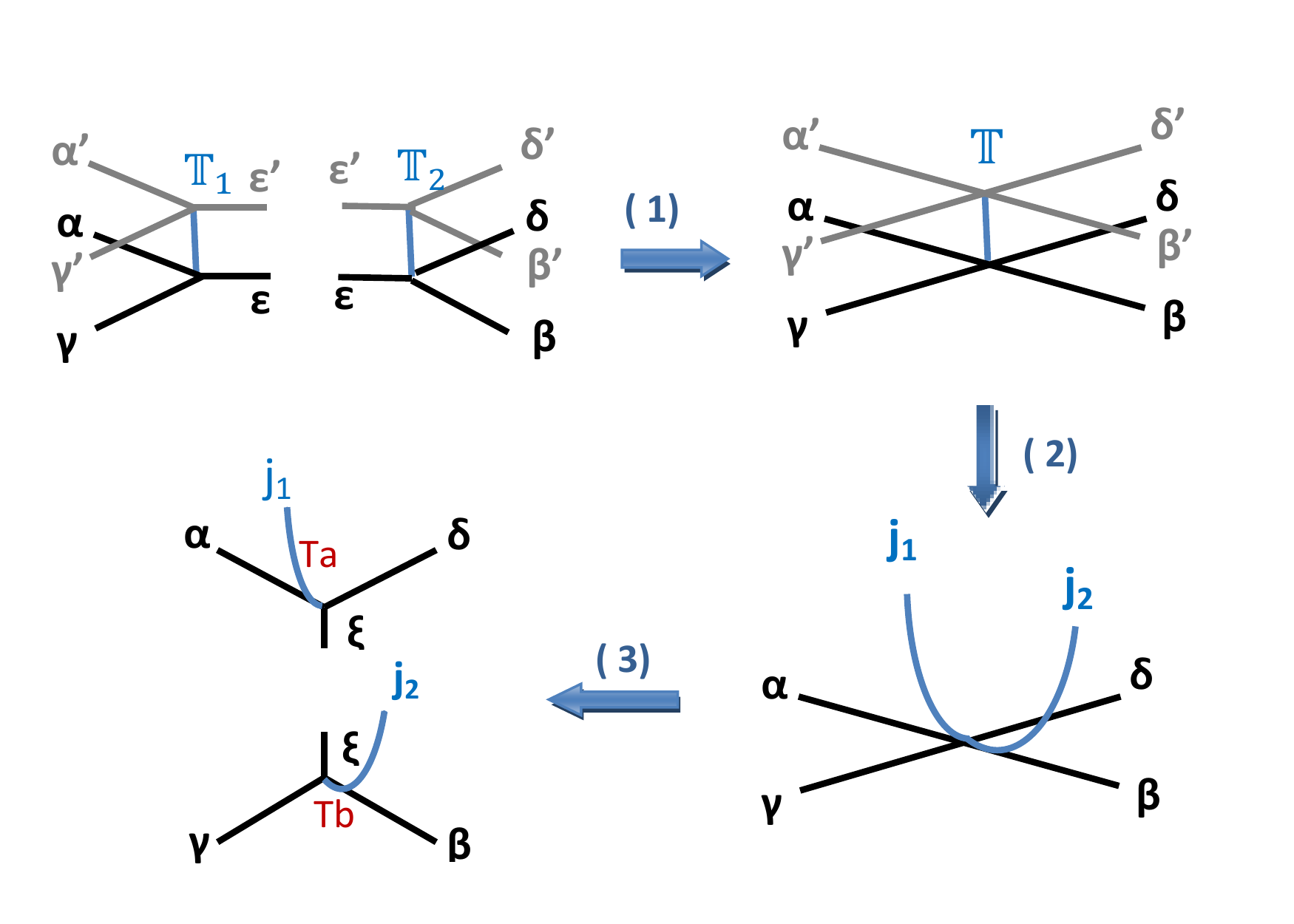,angle=0,width=8cm}} \caption{The
QSRG procedure on honeycomb lattice (part 1): (1) combining
double tensor  $\mathbb{T}_1$  and $\mathbb{T}_2$ on the neighboring
sites into a double tensor $\mathbb{T}$. (2) decomposing
the double tensor $\mathbb{T}$ into the tensor $\tilde{T}$. (3) decomposing the tensor $\tilde{T}$, resulting in tensors
$T_a$, $T_b$. }
\label{o-2DRG-1}
\end{figure}

To implement QSRG, first we form a double tensor $\mathbb{T}$ on each site $\mathbb{T}_{1; \alpha' \gamma'\epsilon',\alpha \gamma\epsilon}=
\sum_i(T^i_{\alpha' \gamma'\epsilon'})^* \times
T^i_{\alpha \gamma\epsilon}$ as shown in Fig. \ref{o-2DRG-1}.
Contract $\mathbb{T}_1$ and $\mathbb{T}_2$ on neighboring sites to form a new double tensor $\mathbb{T}$
\be
\mathbb{T}_{\alpha'\beta'\gamma'\delta',\alpha\beta\gamma\delta}=\sum_{\epsilon,\epsilon'} \mathbb{T}_{1; \alpha' \gamma'\epsilon',\alpha \gamma\epsilon}\mathbb{T}_{2; \epsilon' \beta'\delta',\epsilon \beta\delta}.
\ee
This completes the step (1) of Fig. \ref{o-2DRG-1}. Like in the 1D case, we then spectrally decompose the double tensor,
\begin{align}
\label{doubleTT-2}
\mathbb{T}_{\alpha'\beta'\gamma'\delta',\alpha\beta\gamma\delta}=
\sum_j \lambda_j (\hat{T}^j_{\alpha'\beta'\gamma'\delta'})^*\times
\hat{T}^j_{\alpha\beta\gamma\delta}.
\end{align}
By retaining only the tensors $\hat{T}^j$ with nonzero $\lambda_j$, we apply a local unitary and remove entanglement between the two sites.

One important difference of 2D gapped states from the 1D case is that in 2D the amount of entanglement of a region grows linearly with the boundary of the region. Straight-forward coarse-graining in 2D would lead to unbounded growth in entanglement and an exponential increase in the size of the representing tensor. Therefore a splitting procedure is necessary on $\hat{T}^j_{\alpha'\beta'\gamma'\delta'}$, similar to the Tensor Renormalization Group\cite{trg}, which separates both the inner and the physical indices into two sets.

The inner indices can be naturally separated according to their orientation, for example into $\{\alpha\delta\}$ and $\{\beta\gamma\}$. The best way to split the physical indices, ideally, would be to minimize entanglement in the resulting tensor $\tilde{T}^{j_1j_2}_{\alpha\beta\gamma\delta}$ between $\{j_1\alpha\delta\}$ and $\{j_2\beta\gamma\}$. Such an optimization procedure can be numerically costly to implement. Instead we choose to fix the splitting procedure in a particular way before implementing the QSRG. For example, in \cite{2drg}, one splits $j$ into two sets $\{lm\}$ and $\{mr\}$ by taking the $j$th eigenvector $\hat{T}^j_{\alpha\beta\gamma\delta}$, replacing $\alpha\beta\gamma\delta$ with $lmnr$, and using it as the physical label for the tensor. The total tensor becomes
$\tilde{T}^{\{lm\}\{nr\}}_{\alpha\beta\gamma\delta}=\sum_j\sqrt{\lambda_j}(\hat{T}^j_{lmnr})^*   \times \hat{T}^j_{\alpha\beta\gamma\delta}$
This is a natural and automatic way of splitting the physical index that maintains the structure of the tensor. In our following discussion of Symmetry Protected (SP)-QSRG, however, we need to choose other splitting procedures to also preserve the symmetry of the tensor. Note that the double tensor remains invariant with this splitting.

After the splitting, we do a singular value decomposition of $\tilde{T}$ in the direction orthogonal to the link between $\mathbb{T}_1$  and $\mathbb{T}_2$ and decompose $\tilde{T}$ into $T_a$ and $T_b$ as shown in step (3) of Fig. \ref{o-2DRG-1},
\be
\tilde{T}^{j_1j_2}_{\alpha\beta\gamma\delta}=\sum_{\xi} T^{j_1}_{a;\alpha,\delta\xi} \times T^{j_2}_{b;\beta\gamma\xi}.
\ee

\begin{figure}[ht]
\center{\epsfig{figure=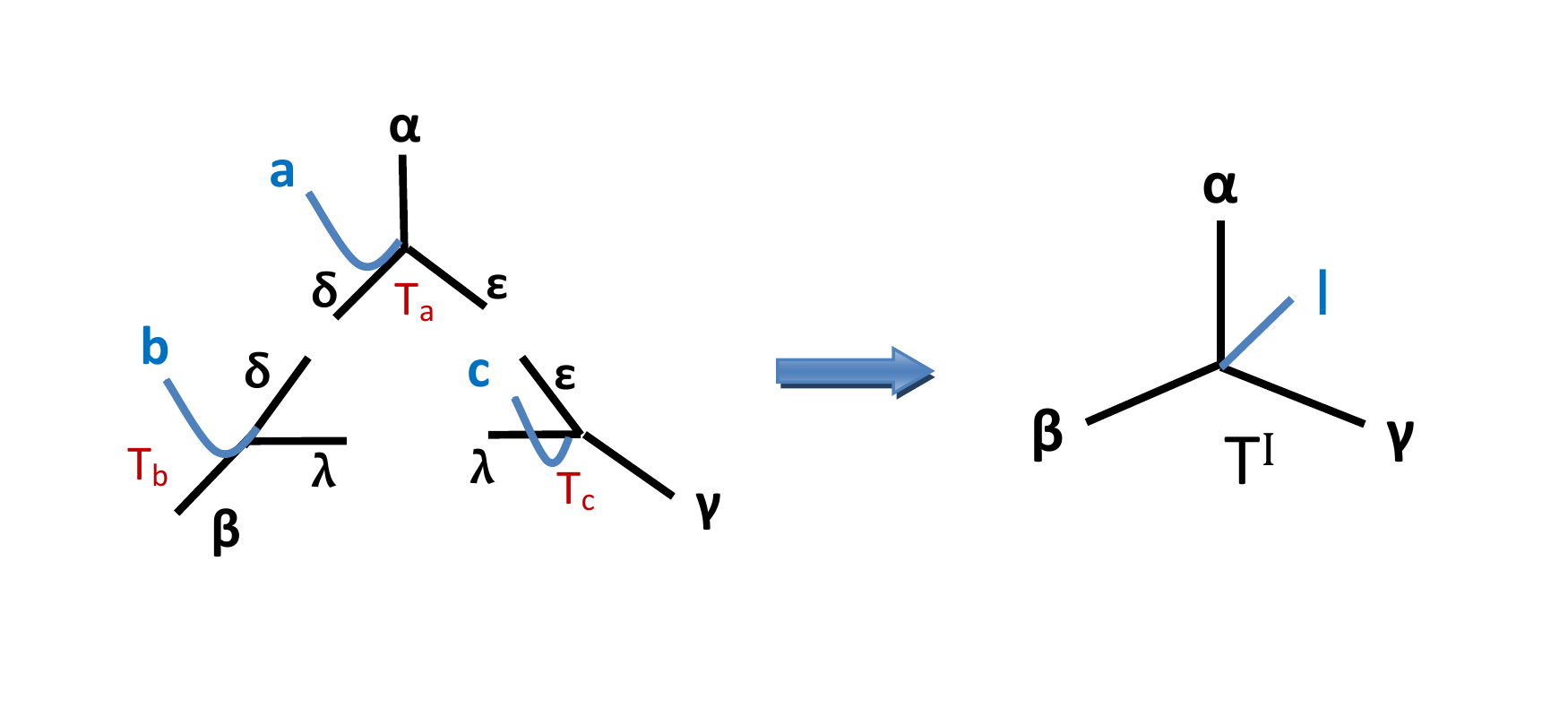,angle=0,width=8cm}} \caption{The
QSRG  procedure on honeycomb lattice (part 2): Merge three tensors
$T_a$, $T_b$ and, $T_c$ around a triangle to form a new tensor
$T^I$.}
   \label{o-2DRG-2}
\end{figure}

Next, we coarse grain the lattice labeled by the tensor $\tilde{T}$
by implementing one step of block decimation in the TRG method
\cite{trg}.  To do this,  we combine the resultant three tensors that meet at a triangle to form a new tensor with physical index $I$  as shown in the Fig.
\ref{o-2DRG-2},
\begin{align}
\label{doubleTT-4}
T^I_{\alpha\beta\gamma}=\sum_{\delta\varepsilon \xi}
T^{a}_{a,\alpha \delta \varepsilon}\times T^{b}_{b,\beta
\delta \xi} \times T^{c}_{c,\gamma \xi \varepsilon}.
\end{align}
where $I$ denotes the combination of all physical indices
$a,b$ and $c$.  This new tensor is the resultant
one after one round of QSRG proposed in \cite{2drg}.

After repeating the above QSRG procedure, the original TPS will then flow into a fixed-point state with SRE removed but not the LRE. Therefore, we can use this method to find out the possible topologically ordered phases by examining the fixed-point tensor.  Moreover, this method is easy to be implemented numerically and has been used to identify and study phases with intrinsic topological order, for example see \cite{jj}. In general, there are certain arbitrariness in choosing the split procedure. Different splitting corresponds to different way to remove local entanglement and could result in different fixed point tensor. However, as long as we follow one particular splitting procedure throughout our RG calculation, tensors within the same phase should flow to the same fixed point form. We will see explicit examples of this in our following discussions.

%%%%%%%%%%%%%%%%%%%%%%%%%%%%%%%%%%%%%%%%%%%%%%%%%%%%%%%%%%%%%%%%%%%%%%%%%%%%%%%%%%%%%%%%%%%%%%%%%%%%%%%%%%%%%%%%%%%%%%%%%%%%%%%%%%%%%%%%%%%%%%%%
%%%%%%%%%%%%%%%%%%%%%%%%%%%%%%%%%%%%%%%%%%%%%%%%%%%%%%%%%%%%%%%%%%%%%%%%%%%%%%%%%%%%%%%%%%%%%%%%%%%%%%%%%%%%%%%%%%%%%%%%%%%%%%%%%%%%%%%%%%%%%%%%

\section{Symmetry protected quantum state renormalization}
\label{SPQSRG}

Although the above QSRG procedure is quite powerful in identifying the intrinsic topological orders, it is not suitable when we consider SPT orders. This is because, the above QSRG procedure is designed to remove short range entanglement, which makes SPT orders indistinguishable from trivial orders. Indeed, consider applying the above QSRG procedure to the 2D dimer state on the honeycomb lattice. Each bond originally has a singlet on it. When applying the steps in Fig.\ref{o-2DRG-1} to the dimer state on a honeycomb lattice, the singlet on the horizontal bond shrinks and, in the most natural and minimum way to re-split in the vertical direction, no singlet is regenerated. After several rounds of RG, it is easy to see that the dimer structure can be completely removed, resulting in a total product state with no SPT order.
\begin{figure}[ht]
\center{\epsfig{figure=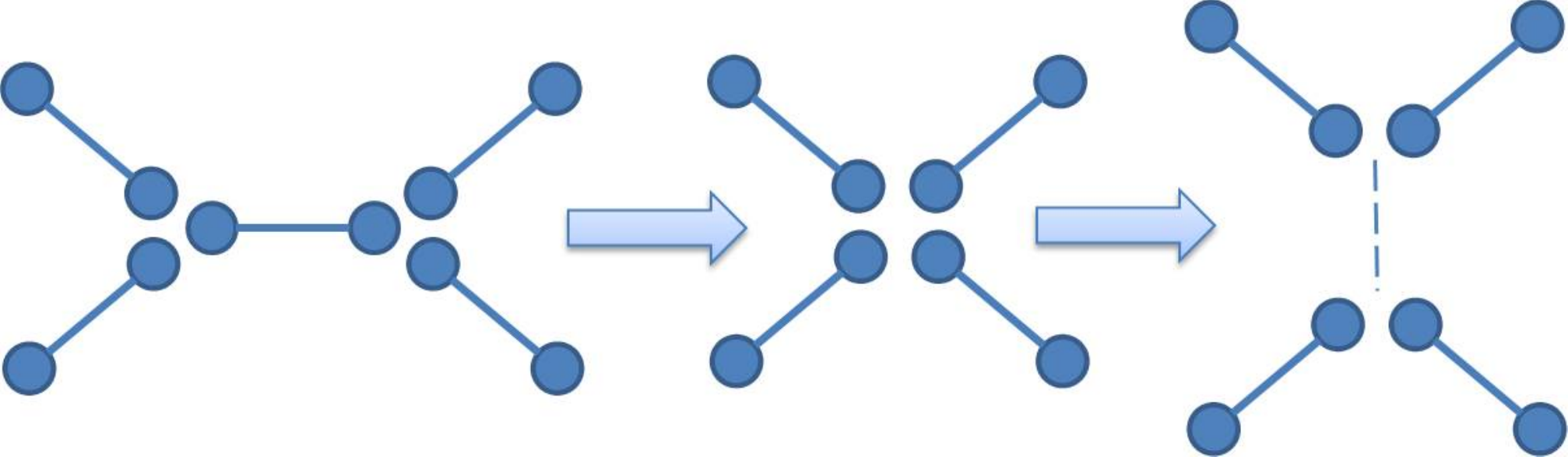,angle=0,width=6cm}}
\caption{Applying QSRG procedure given in \cite{2drg} to
2D dimer state. After applying steps in Fig.\ref{o-2DRG-1}, the
dimer on the middle bond gets removed. The dimer structure can be
completely removed after a couple rounds of RG.}
   \label{QSRG_dimer}
\end{figure}

The key reason that the dimer state flows to a trivial state under the QSRG scheme is because the symmetry of the system is not preserved when doing local unitary transformations on the state. Therefore, we need to find a way to explicitly incorporate the symmetry of the system into the RG scheme in order to preserve the short range entanglement structure of SPT states in the RG flow. In this section, we will propose such a symmetry protected QSRG (SP-QSRG) procedure for TPS and demonstrate its effectiveness in identifying the SPT order from fixed point tensors.

In order to devise the 2D procedure, let's first start with the simpler 1D case. The 1D QSRG procedure as given in \cite{1drg} does preserve the short range entanglement structure in the fixed point tensor of, for example, 1D AKLT state, hence allowing the identification of the SPT order. However, a major difference between 1D and 2D QSRG is that, in 1D the amount of entanglement of a segment of an MPS is bounded, therefore we can perform coarse graining without explicitly removing short range entanglement. In 2D, this is not the case. The amount of entanglement of a region in a TPS grows linearly with the length of its boundary, therefore straightforward coarse graining would lead to unbounded growth of entanglement and hence unbounded growth in the size of the representing tensor. In order to maintain a bounded numerical complexity during the RG procedure, it is important that irrelevant entanglement structures are removed using, for example, the merging and splitting procedure as shown in Fig. \ref{o-2DRG-1}, as discussed in the previous section. The key to our symmetry protected RG scheme is then to preserve symmetry in the merging and splitting procedure. Before we do this in 2D, let's first add the merging and splitting procedure into 1D RG scheme and show how symmetry protection can be incorporated.

An important property of SPT order, which is going to be useful in our following RG procedure, is that it is stable even if addition or removal of local degrees of freedom are allowed as long as they form linear representations of the symmetry group. For example, in the case of Haldane phase with $SO(3)$ rotation symmetry, its edge spin $1/2$ degrees of freedoms are stable even if integer spins can be added locally to the system. The fractional edge spin $1/2$ can become other half integer spin but cannot become a trivial spin $0$ through interaction with any integer spin. In contrast, if we consider a spin $2$  state \footnote{It parent Hamiltonian is different from the AKLT one:  $\sum_i \vec{S}_i \cdot \vec{S}_{i+1} + {1\over 3} (\vec{S}_i \cdot \vec{S}_{i+1})^2$,  but can be constructed in a similar way such that it is in the form of $\sum_{n=1}^4 c_n \sum_i (\vec{S}_i \cdot \vec{S}_{i+1})^n$.} with similar valence bond structure as the spin $1$ AKLT state, then its edge spin $1$  can be removed by adding an extra spin $1$ to the edge which forms a singlet with the original edge spin $1$.
Therefore, the spin $2$ AKLT state is in a trivial SPT phase. Its edge state degeneracy is not protected under $SO(3)$ rotation symmetry. On the other hand, adding or removing projective representation can destroy SPT order and is hence not allowed in the RG flow. Therefore in our SP-QSRG procedure, we are going to add or remove linear, not projective, symmetry representations locally to our need to flow the quantum state to a fixed point without breaking symmetry.

%%%%%%%%%%%%%%%%%%%%%%%%%%%%%%%%%%%%%%%%%%%%%%%%%%%%%%%%%%%%%%%%%%%%%%%%%%%%%%%%%%%%%%%%%%%%%%%%%%%%%%%%%%%%%%%%%%%%%%%%%%%%%%%%%%%%%%%%%%%%%%%%

\subsection{One-dimensional case}
\label{1d}

\subsubsection{Algorithm }

\begin{figure}[ht]
\center{\epsfig{figure=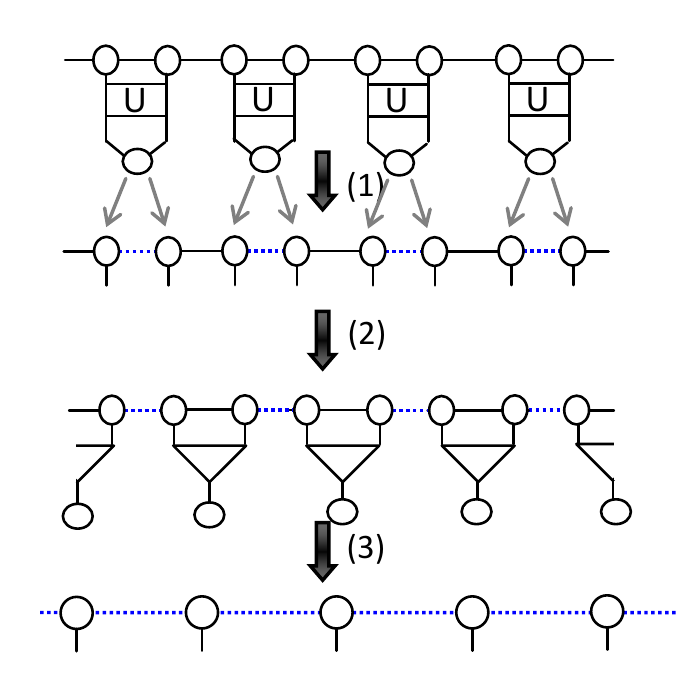,angle=0,width=7cm}}
\caption{Schematic procedure of the revised 1D QSRG. The step (1) is to remove the SRE by performing local unitary transformations. The
steps (2) and (3) are to coarse grain the lattice by removing the
local degrees of freedoms and merging two sites.
}
   \label{1dqsrg-total}
\end{figure}

\begin{figure}[ht]
\center{\epsfig{figure=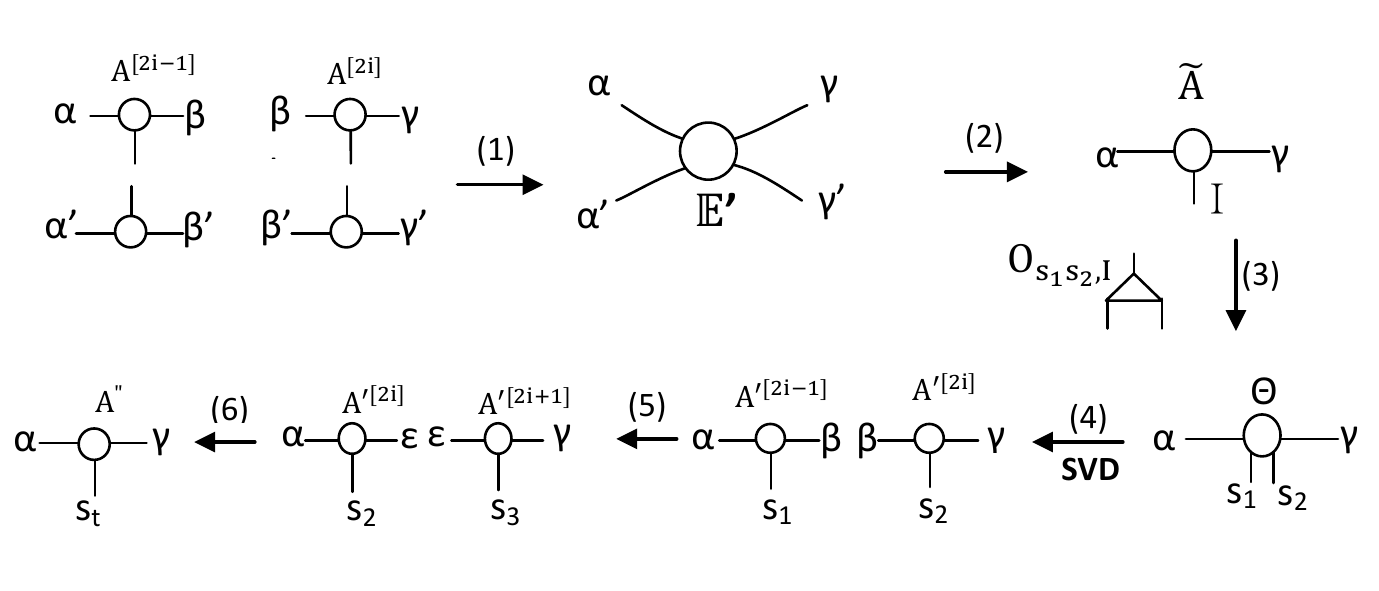,angle=0,width=9cm}}
\caption{Schematic procedure of the SP-QSRG.}
   \label{1DQSRG}
\end{figure}

The schematic procedure of the 1D SP-QSRG is shown in Fig. \ref{1dqsrg-total}. We are going to specify the detailed procedure of splitting and merging of the neighboring site tensors in order to remove the SRE and coarse-graining. We will pay special attention to how the on-site symmetry is preserved in the procedure of splitting and merging. The whole procedure can be broken into two parts: disentangling (see Fig. \ref{1DQSRG} step(1)-(4) and Fig. \ref{1dqsrg-total} step(1)) and coarse-graining (see Fig. \ref{1DQSRG} step(5)-(6) and Fig. \ref{1dqsrg-total} step(3)).

Suppose that the ground state considered is represented as an MPS with a rank-three tensor $A^i_{\alpha,\beta}$, where $i$ is the physical index and $\alpha \;, \beta$ are the left and right indices of the inner bond, respectively. For simplicity of notation, we will assume translational invariance so that the tensor $A$ is the same for all sites.

Our proposed SP-QSRG is implemented as follows.

\begin{enumerate}

\item  Step (1) of Fig. \ref{1DQSRG}:

Firstly, we form a two-site positive double tensor. It is the same as the first step of the original QSRG.

\item Step (2) of Fig. \ref{1DQSRG}:

Then think of $\mathbb{E}'_{\alpha \gamma,\alpha'\gamma'}$ as a matrix with row index $\alpha,\gamma$ and column index $\alpha',\gamma'$. Perform the spectral decomposition as discussed before, i.e.,
\begin{align}
\mathbb{E}'_{\alpha \gamma,\alpha' \gamma'}=\sum_j \lambda_j
(\hat{A}^j_{\alpha',\gamma'})^* \times \hat{A}^j_{\alpha,\gamma}.
\end{align}
where we have kept only the non-zero eigenvalues $\lambda_j$'s and the corresponding eigenvectors $\hat{A}^j_{\alpha,\gamma}$.
With such a spectral decomposition we can form a
new tensor $\tilde{A}^I_{\alpha,\beta}= \sqrt{\lambda_I} \hat{A}^I_{\alpha,\beta} $ whose double tensor is still $\mathbb{E}'$. As mentioned, this step is to perform a local unitary transformation on the two sites, i.e., a disentangler in MERA. In the first two steps of the scheme, symmetry can be naturally preserved.

\item  Step (3)  and (4) of Fig. \ref{1DQSRG}:

Now we add the splitting procedure to the 1D QSRG scheme. While such a procedure is not entirely necessary in 1D, it will be crucial to the 2D QSRG scheme. Split the physical index $I$ into two parts $s_1$ and $s_2$ with a unitary operator $O_{s_1s_2,I}$
\begin{align}\label{step3}
\Theta^{s_1,s_2}_{\alpha,\gamma}=\sum_I  O_{s_1s_2,I} \times\tilde{A}^I_{\alpha,\beta}
\end{align}
Then perform a singular value decomposition and split $\Theta^{s_1,s_2}_{\alpha,\gamma}$ into two tensors $A'^{s_1}_{i-1}$ and $A'^{s_2}_{i}$
\be
\Theta^{s_1,s_2}_{\alpha,\gamma}=\sum_{\beta} (A'^{s_1}_{\alpha\beta})^{[2i-1]}(A'^{s_2}_{\beta\gamma})^{[2i]}
\label{step4}
\ee
In these two steps, symmetry needs to be carefully preserved as we discuss below.

Step (3), as given in Eq.\ref{step3}, involves locally adding degrees of freedom. From the discussion about SPT phases in section \ref{SPT}, we see that in order to preserve SPT order, only linear symmetry representations can be added or removed in the RG process of a system. In practice, this will mean that the physical indices of the site tensors should carry the same class of projective representation before and after these steps. Therefore, when we split the site tensor $\tilde{A}$, we should make sure that $s_1$ and $s_2$ are in the same class of projective representations as the physical index of the original tensor $A$. If necessary, we should enlarge the Hilbert space of the physical index of $\tilde{A}$ when splitting. On the other hand, if the symmetry representations on each site can change during the RG process, the SPT order of the state can be totally lost. To do the splitting, we need to use an operator $O_{s_1s_2,I}$. We will discuss how to determine this operator $O$ in the next section.

For example, if the on-site symmetry protecting the SPT is $SO(3)$, then the representations for the integer spins and the half-integer spins are in different projective classes. Therefore, if we start with an integer spin chain model, we should make sure that after merging and splitting, the physical indices of each site are still integer spins. By `measuring' the quantum numbers of the original physical index, we can properly assign the symmetry representation to the new physical indices. We demonstrate an explicit procedure with the AKLT example in the next section.

Similar procedure can also be followed for other on-site symmetry groups such as $Z_2 \times Z_2$. In such a case, irreducible linear representations are all one dimensional and irreducible projective representations are all two dimensional. When splitting the tensor $\tilde{A}$ we should make sure that the physical indices of the resultant tensors $A'$'s are in the same class of representations as that of $A$.

Symmetry preserving considerations are also necessary in step (4). In step (4), when doing the singular value decomposition as described in Eq. \ref{step4}, a cut-off might be necessary on the dimension of $\beta$ to keep the computation efficient. When doing the cut-off, we need to preserve the symmetry structure of the tensor by keeping or removing degenerate blocks together. If we perform the cut-off without respecting the block structure, systematic information about the symmetry of the system can be lost. Such symmetry considerations apply similarly to all groups.

\item Step (5) and (6) of Fig. \ref{1DQSRG}:

We then merge alternatively the $A'$ tensor on site $2i$ and $2i+1$ to form a new site tensor $A''$, i.e.,
\begin{align}\label{T''}
A''^{s_t}_{\alpha,\gamma}=\sum_\epsilon  (A'^{s_2}_{\alpha,\epsilon})^{[2i]}
\times (A'^{s_3}_{\epsilon,\gamma})^{[2i+1]}.
\end{align}
Here, $s_t$ denotes the conbination of all physical indices $s_2$ and $s_3$.

\end{enumerate}

This will complete one round of SP-QSRG for 1D MPS. Schematically the whole procedure on the tensors is shown in Fig. \ref{1DQSRG} and the corresponding renormalization operation on the quantum state is shown in Fig.\ref{1dqsrg-total}. The key to this procedure is in preserving the symmetry of the state, which we demonstrate with the example of the AKLT state in the next section.

%%%%%%%%%%%%%%%%%%%%%%%%%%%%%%%%%%%%%%%%%%%%%%%%%%%%%%%%%%%%%%%%%%%%%%%%%%%%%%%%%%%%%%%%%%%%%%%%%%%%%%%%%%%%%%%%%%%%%%%%%%%%%%%%%%%%%%%%%%%%%%%%

\subsubsection{1-dimensional AKLT state as an example }

We now demonstrate how to apply the SP-QSRG procedure discussed above to a concrete example of an SPT state. Let us consider the 1D AKLT \cite{AKLT1987} MPS with $A^i=\sigma^i$, where $\sigma^i$ $ (i=x,y,z)$ are the Pauli matrices. This AKLT state is shown to have SPT \cite{1daklt0,1daklt1,1daklt2} order. Moreover, as shown in \cite{1drg}, the fixed-point state of the original QSRG is the dimer state, a product of singlets between each pair of neighboring sites, characterized by the transfer matrix $\sum_{i,j=1}^2 \frac{1}{4}|ii\rangle\langle jj|$ obtained from its MPS representation. As we mentioned, in 1D this QSRG does not destroy the symmetry and so preserves the SPT order. Thus, the dimer and AKLT state have the same SPT order and we expect them to flow to the same fixed point under our SP-QSRG procedure as well. We want to emphasize again that while the SP-QSRG procedure is not necessary to detect SPT order in 1D, we study it here in preparation for the 2D case where the original QSRG procedure fails to preserve the SPT structure.

\begin{figure}[ht]
\center{\epsfig{figure=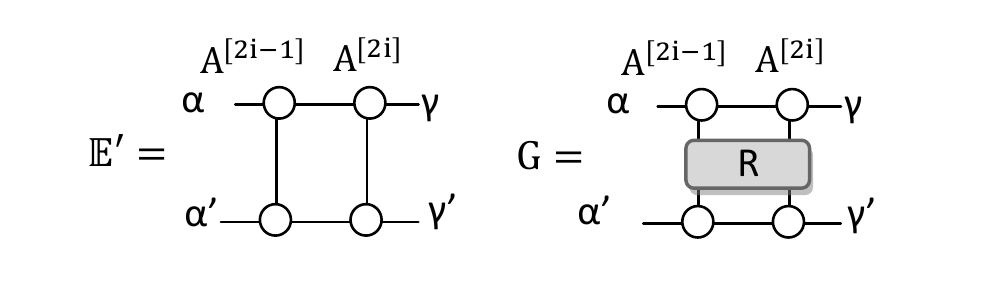,angle=0,width=7cm}}
\caption{ (Left): The two sites transfer matrix $\mathbb{E}'$. (Right): The two sites generalized transfer matrix $G$ by inserting an operator $R$.   }
   \label{insert_operator}
\end{figure}

We now apply our SP-QSRG to the AKLT MPS. We first apply the disentangler, and then perform the coarse-graining step as discussed in the previous section. When doing spectral decomposition of the double tensor (the step (2) of Fig. \ref{1DQSRG}), we find that the eigenspectrum of the double tensor $\mathbb{E}'$ has one- and three-fold degeneracies. This reflects the $SO(3)$ symmetry of the AKLT state. In order to manipulate the physical index without breaking this $SO(3)$ symmetry, we need to identify the spin representations of the physical index $I$ by measuring the $(\vec{S})^2$ and $S_z$ quantum number. To do this, we can insert operators $R=\mathbb{I}+\epsilon S_z$ and $R=\mathbb{I}+\epsilon (\vec{S})^2$, $\epsilon$ being a small number, into the transfer matrix $\mathbb{E}'$ to form  generalized transfer matrices $G$ (see Fig. \ref{insert_operator}). With this insersion, the degeneracy pattern of $\mathbb{E}'$ will be split. By comparing the eigenvalue of $\mathbb{E}'$ and $G$ for the same eigenvector, we can read off the quantum number.

After determining the spin representation on two sites as a direct sum of spin $0$ and spin $1$, we can append a spin $2$ sector (with zero weight) and decompose the Hilbert space into two spin $1$'s, i.e., $3 \otimes 3=1 \oplus 3 \oplus 5$. This decomposition is realized by the following operator $O_{s_1s_2,I}$.
\begin{align}\label{O1} \notag
O_{s_1s_2,I} = &\frac{1}{\sqrt{3}} (|1,-1\rangle-|0,0 \rangle + |-1,1 \rangle ) \langle J=0,J_z=0|+\\  \notag
               &\frac{1}{\sqrt{2}} (|1,0 \rangle-|0,1\rangle   ) \langle J=1,J_z=1|+ \\ \notag
               &\frac{1}{\sqrt{2}} (|1,-1\rangle-|-1,1\rangle) \langle J=1,J_z=0|+ \\
               &\frac{1}{\sqrt{2}} (|0,-1 \rangle-|-1,0\rangle) \langle J=1,J_z=-1|,
\end{align}
where $|s_1,s_2\rangle$'s with $s_{1,2}=-1,0,1$ are the basis vectors of two spin $1$. $J$ and $J_z$ are the total spin quantum numbers. $I$ is the physical index after doing spectral decomposition of $\mathbb{E}$, and $s_1, s_2$ are new physical indices in the  integer spin representation. Here, the $O_{s_1s_2,I}$ is a unitary operator and corresponds to the Clebsch-Gordan coefficients. The total spin $2$ sector is omitted as it has $0$ weight.

We then identify the labels $s_{1,2}$ as the physical indices of the two new site tensors $A'$, respectively and perform the SVD decomposition as shown in step (4) of Fig.\ref{1DQSRG}. In this decomposition, we ensure that the physical spin on the new sites also have integer-spin representation and we have split the tensor $\tilde{A}$ to $A'$'s without destroying the SPT. We then form the new tensor $A''$ to complete one round of our SP-QSRG.

In this way, we obtain the fixed point double tensor $\mathbb{E}_{\alpha\beta,\alpha' \beta'}$ of AKLT state keeping $\chi=2$ as follows:
\begin{align}\label{aklt1} \notag
 & \mathbb{E}_{11,11}=0.524, \qquad  \mathbb{E}_{11,22}=0.053,\\
 \notag & \mathbb{E}_{12,12}=0.471, \qquad  \mathbb{E}_{21,21}=0.471,\\
 & \mathbb{E}_{22,11}=0.053, \qquad  \mathbb{E}_{22,22}=0.524.
\end{align}
We also consider the dimer state, whose MPS can be written as
\begin{align}\label{dimer}\notag
&A_{00}^{01}=-1, \qquad  A_{01}^{00}= 1,  \\
&A_{10}^{11}=-1, \qquad  A_{11}^{10}= 1.
\end{align}
This phase also has $SO(3)$ symmetry, and it is obvious that the eigenvalues of  $\mathbb{E}'$ has the $1\oplus 3$ pattern. Hence, we can use the same splitting procedure, as mentioned above, to the dimer state. The fixed point tensor for the dimer state is the same as the one for the AKLT state.  It then justifies our SP-QSRG in identifying the SPT order.

The first point to note is that this fixed point tensor seems to have a complicated form. The corresponding fixed point state has a nonzero correlation length and is different from the dimer state which one might have expected. This complication is due to the specific splitting procedure $3 \otimes 3=1 \oplus 3 \oplus 5$ we are using. Short range entanglement is not removed in an optimal way using this splitting procedure, resulting in a fixed point state with residue correlation and entanglement. Even so, we are still able to see that the AKLT and the dimer states have the same SPT order by flowing them to the same (non-ideal) fixed point form. By knowing the SPT order of the dimer state, we can determine the SPT order of the AKLT state. There is a further point which needs to be clarified.  The fixed point tensors might be not unique owing to the scale and the basis transformations of inner and physical indices. However, we can do a normalization to fix the scale, and remove the basis ambiguity of the physical index by forming the double tensor. In this particular case, basis transformations of the inner indices do not change the double tensor as they corresponds to spin rotation operations of the virtual spin $1/2$'s and are symmetries of the double tensor. Therefore, we obtain a unique form of the double tensor. In more general situations (with $\chi>2$), basis transformation of the inner indices can change the double tensor and we need to construct quantities invariant under such changes to distinguish different fixed point tensors.
% The gauge transformation of inner indices have been fixed specially, because the projective representations of inner induces correspond to different symmetric sectors for $\chi=2$. When splitting the tensor $\tilde{A}$ to $A'$'s, the singular value spectrum is double degeneracy. Then, the corresponding vectors will be fixed owing to different symmetric sectors.

% another way
The points made so far apply in principle to any other splitting procedures which may be used and can lead to different fixed point tensors. Here we present a procedure with a different operator $O'$ which actually has the dimer state as the fixed point state. Ideally, one would want to optimize the splitting procedure to get simpler fixed point tensors. However, as the optimization procedure is difficult to implement, we retain the arbitrariness in the splitting procedure and rely on the ability of the algorithm to flow states within the same SPT phase to the same fixed point to identify SPT order.

We can  perform the unitary operator $O'$ of the SP-QSRG in another way by enlarging the Hilbert space of the physical indices to contain 4 spin $1/2$'s. The unitary operator can be written as
\begin{align}\label{O2}\notag
O'_{s_1s_2,I} = & \frac{1}{2}(|01\rangle-|10\rangle)_{14}(|01\rangle-|10\rangle)_{23} \langle J=0,J_z=0|+\\  \notag
                & \frac{1}{\sqrt{2}}|00\rangle_{14}(|01\rangle-|10\rangle)_{23} \langle J=1,J_z=1|+ \\ \notag
                & \frac{1}{2}(|01\rangle+|10\rangle)_{14}(|01\rangle-|10\rangle)_{23} \langle J=1,J_z=0|+ \\
                & \frac{1}{\sqrt{2}}|11\rangle_{14}(|01\rangle-|10\rangle)_{23} \langle J=1,J_z=-1|,
\end{align}
where spins labelled by $1$ and $2$ belong to $s_1$ and the ones by $3$ and $4$ belong to $s_2$. Spins labelled by $2$ and $3$ always form a singlet, and the ones by $1$ and $4$ form a spin $0$ or a spin $1$ corresponding to the total $J$.
$s_{1,2}$ both form integer spin representations. So the spin representation of the physical index per site is kept with this $O'$ operator. Using this splitting procedure, we will obtain the following fixed-point double tensor for the AKLT state:
\begin{align}\label{aklt2}\notag
 & \mathbb{E}_{11,11}=0.5,\qquad  \mathbb{E}_{12,12}=0.5,\\
  & \mathbb{E}_{21,21}=0.5,\qquad  \mathbb{E}_{22,22}=0.5.
\end{align}
Again, this is the same as the one for the dimer state obtained by the same SP-QSRG procedure. It is easy to show that this fixed point tensor can be obtained from the MPS  \eq{dimer}, therefore the dimer state is actually the fixed point state of this SP-QSRG procedure. Note that the resultant double tensors in \eq{aklt1} and \eq{aklt2} are different. This reflects the fact that different SREs are removed during the process of QSRG. But as along as we use the same procedure in the RG process, the same fixed point can be reached.

To take another example, this degeneracy pattern also could be interpreted as combining two spin $1/2$ sites, i.e., $2 \otimes 2=1\oplus 3$. If we take this interpretation and perform the further splitting to obtain two $A'$ tensors. Then, the physical index of $A'$ will be in spin $1/2$ representation. We will obtain the the fixed point double tensor, $\mathbb{E}_{11,11}=1$, which is a product state with no entanglement. This is in the wrong class of projective representation.  These above examples make it clear that if we ensure the projective representation in the same class in the QSRG process, the SPT phase will flow to nontrival fixed point.

\subsubsection{1-dimensional toy model }

In above example, we start from a MPS and flow it to fixed point by using SP-QSRG procedure. Here, we show you that this algorithm can also be used to study the phase transition between different SPT phases. The model we study is the staggered Heisenberg model on the 1D spin $1/2$ chain. It is a quantum antiferromagnetic Heisenberg model with alternating strength in the nearest-neighbor-exchange couplings, and its Hamiltonian is
\begin{align}\label{1dchain}
H= \sum_i J_1 \vec{\sigma}_{a_i} \cdot \vec{\sigma}_{b_i}+ J_2 \vec{\sigma}_{b_i} \cdot \vec{\sigma}_{a_{i+1}}.
\end{align}
Here, each site contains two spin $1/2$ particles. $J_1$ bond (the thin one) means the interaction between two inner particles and $J_2$ bond (the thick one) means the interaction between two nearest-neighbor sites. This model is $SO(3)$ invariant.

\begin{figure}[ht]
\center{\epsfig{figure=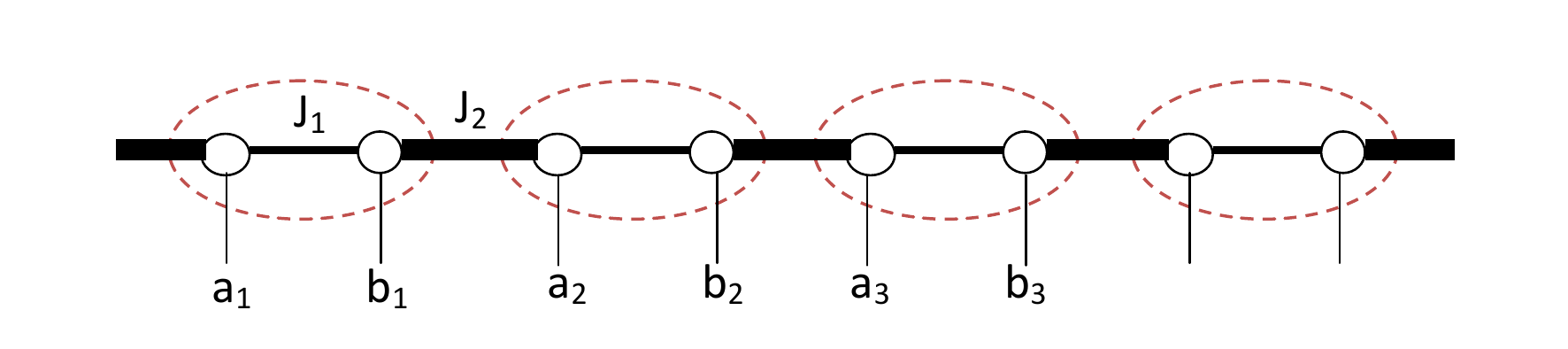,angle=0,width=8cm}}
\caption{ The spin chain model on the 1D spin chain with two different nearest-neighboring bond couplings $J_1$ and $J_2$ (thin and thick, respectively). Each site (dashed ellipse) contains two spin $1/2$ particles,  labeled as $a_i$ and $b_i$. }
   \label{1dmodel}
\end{figure}

By using the iTEBD with $\chi=12$, we can find the ground states of the Hamiltonian \eq{1dchain}, with $J_1=1-g$ and $J_2=g$, $0 \leq g \leq 1$. At $g=0$, the ground state is a trivial product state of on-site spin $0$'s. At $g=1$, the ground state is a dimer state and has nontrivial SPT order. At some critical value of $g$, the state must go through a phase transition. One way to detect the phase transition is to apply our RG procedure. We find that at $g>0.5$, the ground state flows to the non-trivial fixed point while $g<0.5$ the ground state flows to a trivial fixed point. At $g=0.5$, the system is gapless, and it is hard to describe this state with finite bond dimension $\chi$ of MPS. Here we use the splitting procedure given in \eq{O2}.

\begin{figure}[ht]
\center{\epsfig{figure=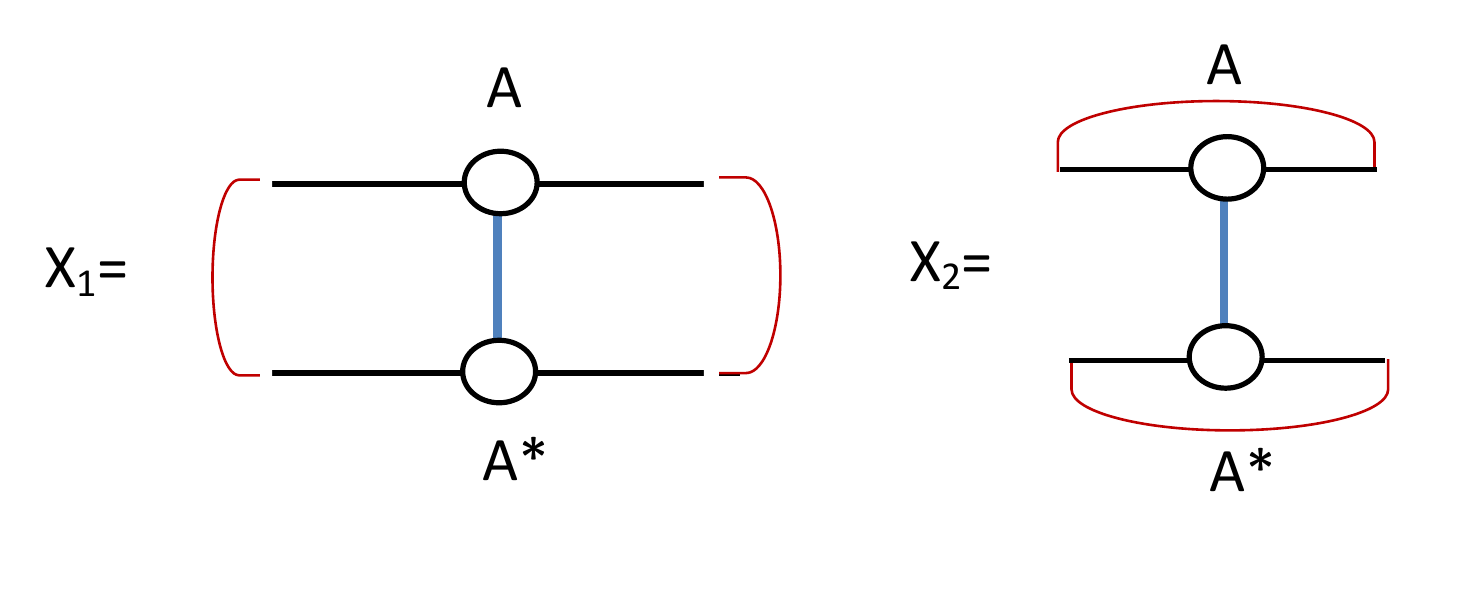,angle=0,width=5cm}}
\caption{ Quality $X_2/X_1$ obtained by taking the ratio of the contraction value of the 1D double tensor in two different ways. }
   \label{1D-x1x2}
\end{figure}

\begin{figure}[ht]
\center{\epsfig{figure=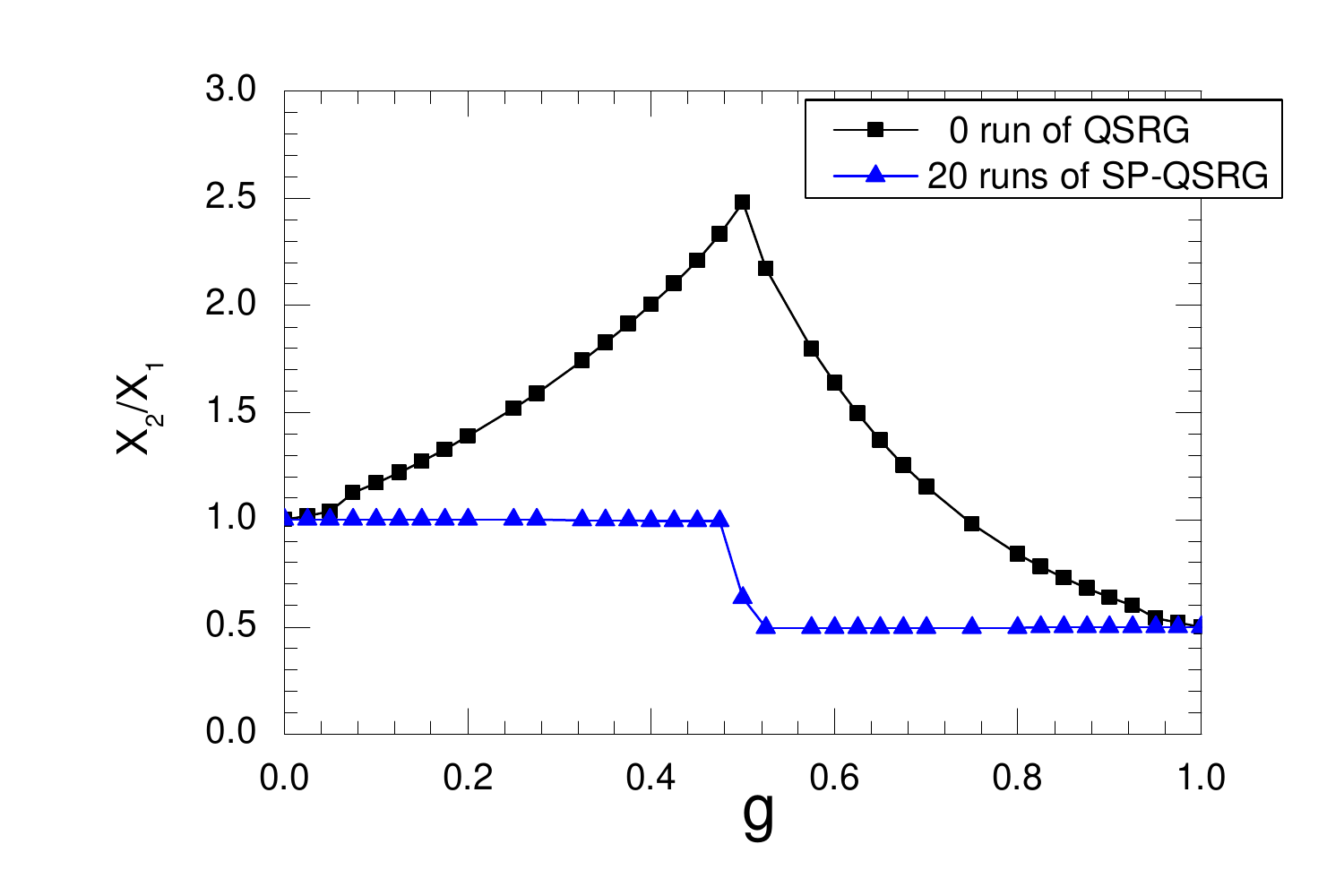,angle=0,width=7cm}}
\caption{ Quality $X_2/X_1$ as a function of parameter $g$. The more runs of the SP-QSRG go, the
more $X_2/X_1$ approaches a step function across phase transition.}
   \label{1dtoy}
\end{figure}

The fixed point tensor may not have a unique form due to the change in scale, basis transformation on the inner / physical dimensions, etc. In order to remove the influence of these factors, we can calculate the basis and scale independent quantity $X_2/X_1$ to distinguish these two fixed points. Such quantity is defined as
\begin{align}\label{eq:X1X2}\notag
&X_1= \sum_{s,i,j} A^s_{i,j} \times  (A^s_{i,j})^* \\
&X_2= \sum_{s,i,j} A^s_{i,i} \times  (A^s_{j,j})^*
\end{align}
as shown in Fig. \ref{1D-x1x2}.  With RG procedure \eq{O2}, we flow the ground states for arbitrary $g$ to fixed points and calculate $X_2/X_1$.  The results are plotted in Fig. \ref{1dtoy}.  As the number of renormalization steps increases, the transition approaches a step function.

%%%%%%%%%%%%%%%%%%%%%%%%%%%%%%%%%%%%%%%%%%%%%%%%%%%%%%%%%%%%%%%%%%%%%%%%%%%%%%%%%%%%%%%%%%%%%%%%%%%%%%%%%%%%%%%%%%%%%%%%%%%%%%%%%%%%%%%%%%%%%%%%

\subsection{Two-dimensional case}
\label{2d}

We will now generalize our SP-QSRG to 2D.  We first describe the algorithm for our SP-QSRG on the honeycomb lattice. Then, we study a deformed 2D AKLT model on the honeycomb lattice, and show that there exists a possible SPT phase by calculating some order parameters. Finally, we demonstrate the power of our SP-QSRG by applying it to this AKLT phase and identify the SPT order.

\begin{figure}[ht]
\center{\epsfig{figure=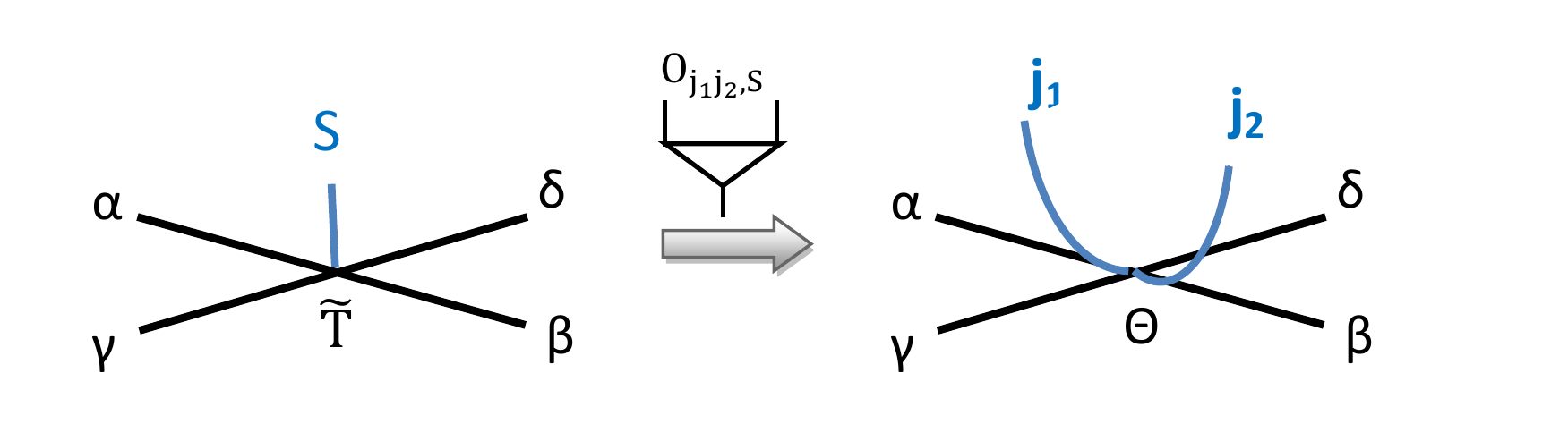,angle=0,width=8cm}}
\caption{ The
SP-QSRG procedure on honeycomb lattice. Enlarging the Hilbert space for the physical indices of the tensor $\tilde{T}$ to ensure the physical indices $j_1$ and $j_2$ are in the same class of projective representation as the original site physical index.}
   \label{sym-2DRG}
\end{figure}

The symmetry protected version of the 2D QSRG scheme contains the same steps as that shown in Fig. \ref{o-2DRG-1}. The key difference is to preserve
symmetry in the merging and splitting procedure.
As for the 1D case, the SP-QSRG in 2D contains two parts: disentangling (see Fig. \ref{o-2DRG-1}) and coarse-graining (see Fig. \ref{o-2DRG-2}). In the disentangling step,  we should be careful about the projective representations of the physical indices when splitting the tensor $\tilde{T}$ in Fig. \ref{sym-2DRG} before re-merging.  We may need to enlarge the Hilbert space for the physical indices of the tensor $\tilde{T}$ to form the tensor $\Theta$ before splitting and ensure that $j_1$ and $j_2$ are in the same class of projective representation as the physical indices on the original sites. After splitting, we merge the three new tensors around a vertex as shown in Fig. \ref{o-2DRG-2}. This completes one round of the SP-QSRG for the weak SPT phases. Moreover, in 2D QSRG, we need to truncate the minor singular values in step (3) of Fig.\ref{o-2DRG-1} to make the numerical computation viable. Of course, the truncation should be performed in accordance with the degeneracy patterns of the singular values to ensure the symmetry property of the SPT.

Again, for $SO(3)$ on-site symmetry the integer spins and half-integer spins belong to different classes of projective representations. Therefore, we should assure that the physical indices of each site tensor in the process of QSRG are in the same class of the spin representation to keep the SPT order. For 1D SPT states with $SO(3)$ symmetry, physical indices on each site always have to be integer spins. However for 2D SPT states with $SO(3)$ and translation symmetry, physical states per each site can be either integer spins or half integer spins, depending on the lattice. The key to preserving the weak SPT order in the 2D QSRG scheme is to keep the spin representation class of the physical index on each site.  Note that we are keeping only block translation symmetry (with odd block size) in the RG procedure which is sufficient to preserve the weak SPT order.

\subsubsection{2-dimensional AKLT phase as the example }

To demonstrate the power of our SP-QSRG, we consider a 2D model with weak SPT order. The simplest one is the 2D AKLT model \cite{AKLT1988} on the honeycomb lattice, which has $SO(3)$ on-site symmetry and translation symmetry. Its ground state -- the AKLT state has a simple TPS representation, see \cite{2daklt,2dMiyake2011} for example.

To be more general, we consider the simple variation of the AKLT model with following Hamiltonian for spin $3/2$ per site on the honeycomb lattice,
\begin{align}\label{2dH}
H= \sum_{<ij>}[ J_1 \vec{S_i}\cdot \vec{S_j}+ J_2( \vec{S_i}\cdot
\vec{S_j})^2+ J_3( \vec{S_i}\cdot \vec{S_j})^3 ]
\end{align}
Obviously, this model has $SO(3)$ on-site symmetry and translation symmetry. In this section, we use iTEBD method to find the ground states of Hamiltonian \eq{2dH} and calculate the expectation values of $M^z_s$ and $M^z$. We identify a region where both order parameters are zero which could potentially be an SPT phase. Then, using SP-QSRG, we flow the ground states to fixed points and identify the order. This model corresponds to the AKLT one if  $J_1=1$, $J_2=116/243$ and $J_3=16/243$. In such a case, its ground state has a closed form in terms of TPS \cite{2daklt} with each on-site spin being decomposed into three spin $1/2$ virtual particles, as shown in Fig. \ref{VBS}.

\begin{figure}[ht]
\center{\epsfig{figure=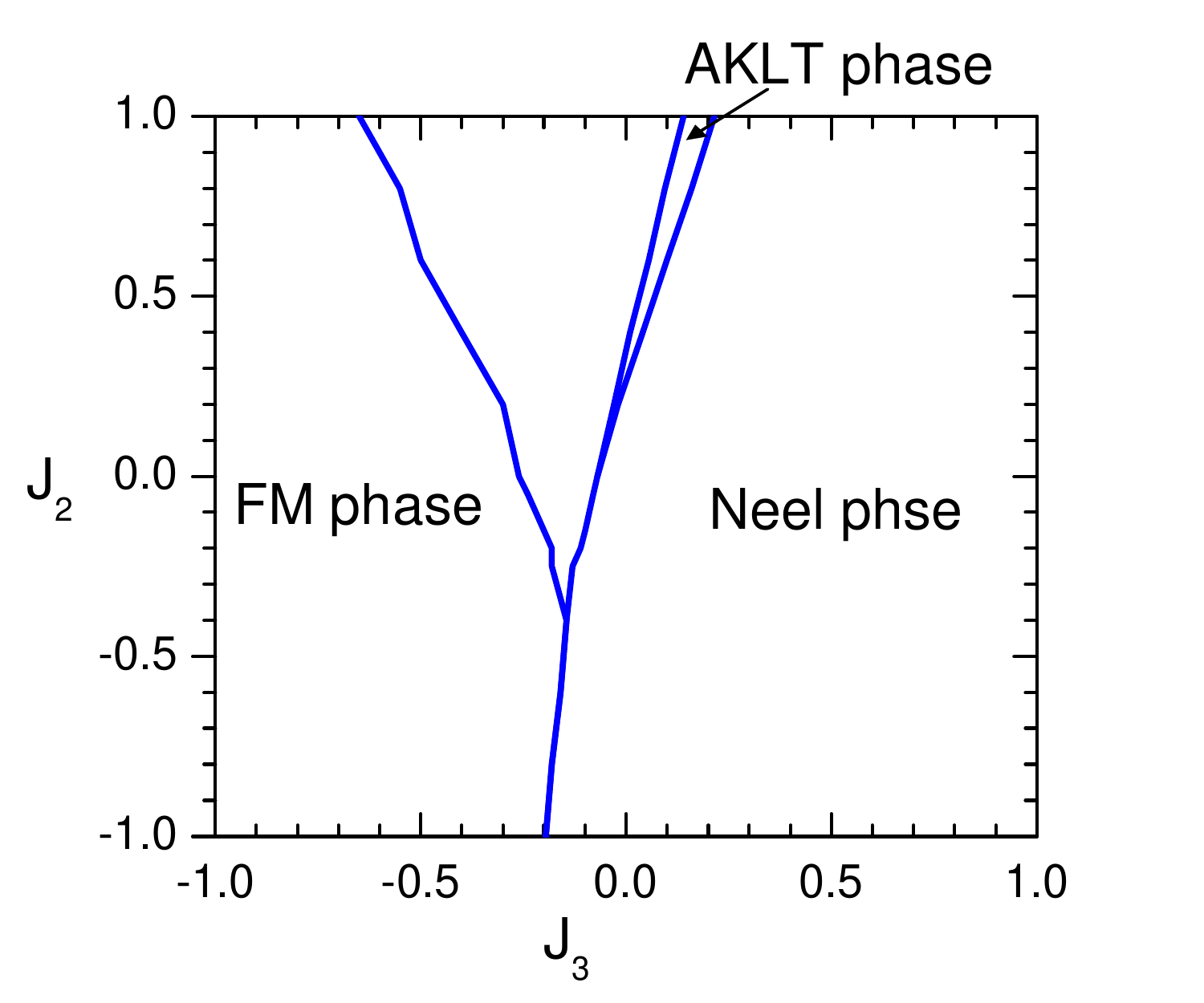,angle=0,width=8cm}}
\caption{The
phase diagram of the deformed AKLT-like model \eq{2dH} by tuning
$J_2$ and $J_3$ with $J_1=1$ fixed. The labels Neel, FM and AKLT
stand for the Neel, the ferromagnetic and the AKLT phases,
respectively. } \label{phase diagram}
\end{figure}

However, for generic values of $J_1$ and $J_2$ there is no known analytic ground state solution of Hamiltonian \eq{2dH}. Instead, we solve this model numerically by using the method of simple update \cite{itebd} for the ground state. This method is to numerically evolve the ansatz state in imaginary time with the help of Trotter decomposition when updating the TPS by each 2-site term of \eq{2dH}. We then apply the method of Tensor RG \cite{trg} to calculate the order parameters, from which we deduce the general structure of the phase diagram.
%One then reduce a finite lattice into a single small cell containing the defects so that the expectation value of physical observables located at the defects is equivalent to the one for the reduced cell.  This method bypass the exponential growth with the size of the TPS when contracting the inner and physical indices to evaluate the expectation values. Though the above method cannot yield very accurate phase diagram, especially near the quantum critical points, we are satisfied with its accuracy when only topological quantities like SPT order are concerned.

\begin{figure}[ht]
\center{\epsfig{figure=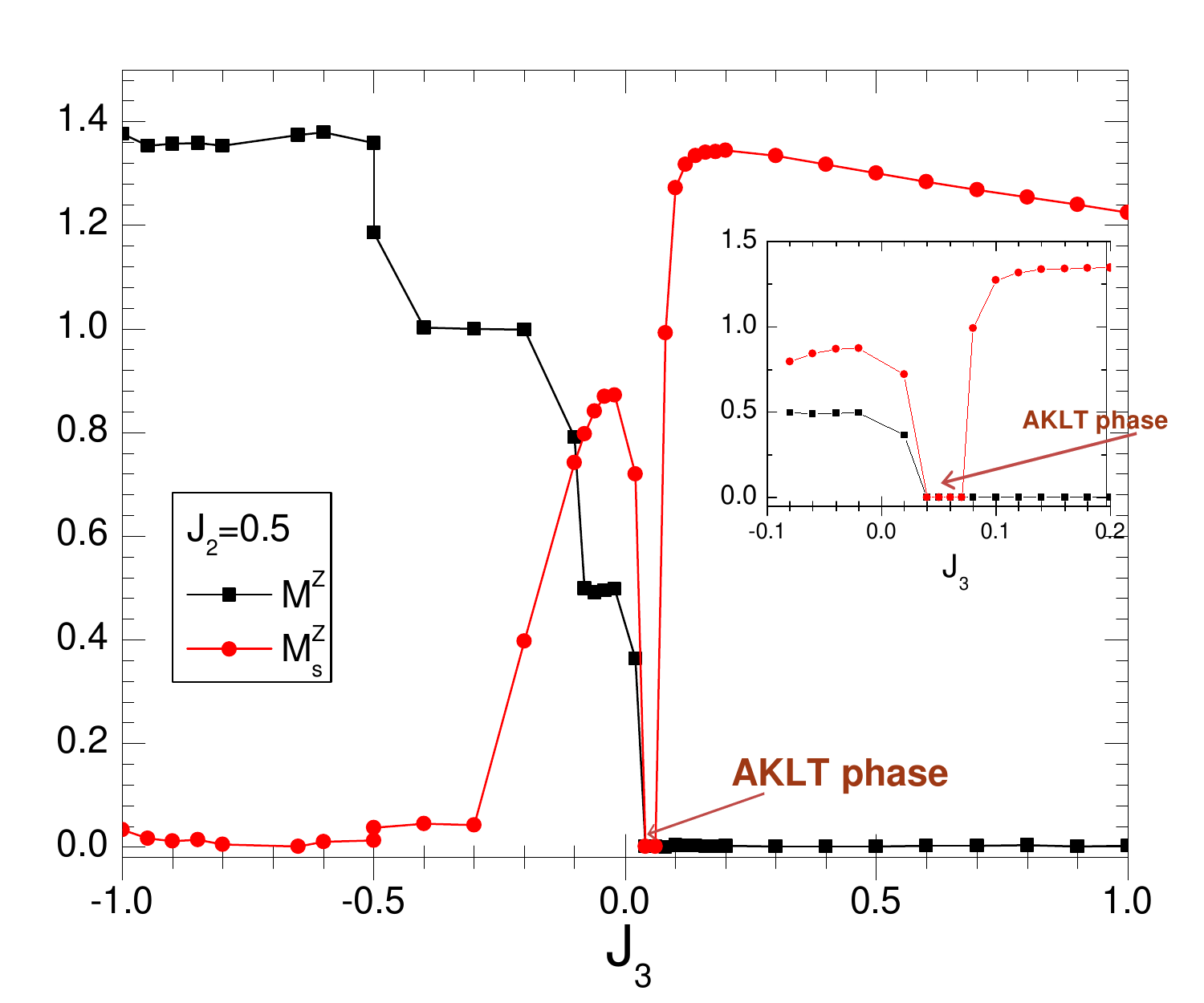,angle=0,width=8cm}}
\caption{A
typical phase diagram for the magnetization $M^z$ and the staggered
magnetization $M^z_s$ by tuning $J_3$ with $J_1=1$ and $J_2=0.5$
fixed.  We fix the bond dimension $\chi=4$ and $D_{cut}=24$ in this
numerical calculation.  }
\label{J20.5}
\end{figure}

After a tedious procedure of scanning the parameter space of the Hamiltonian \eq{2dH} for the magnetization $M^z$ and the staggered magnetization $M^z_s$, we finally obtain its total phase diagram as shown in Fig. \ref{phase diagram}. If $M^z_s=0$ and $M^z\ne 0$ we then identify it as in the ferromagnetic (FM) phase. On the other hand, if $M^z_s\ne 0$ but $M^z = 0$, we identify it as in the Neel phase. In a region where both order parameters are non-zero, we cannot identify the phase and the system is probably gapless. For more discussions on numerical results, see Appendix \ref{app b}.

From our numerical results, we find there is a regime in the parameter phase in which neither $M^z_s$ nor $M^z$ is nonzero. For example Fig.\ref{J20.5} shows the $M^z$ and $M^z_s$ value with $J_1=1$, $J_2=0.5$ and $J_3=-1.0$ to $1.0$ in which the order parameters goes to zero near $J_3=0$.  On the other hand, as the system does not break the on-site $SO(3)$ and translational symmetry, it could be in a weak SPT phase. If this is true, we will call it the AKLT phase. Now, we will apply our SP-QSRG to give the supporting evidence. It turns out that this is a SPT phase.

Our examination based on QSRG procedure goes as follows.
First, we apply the usual QSRG to the ground state of \eq{2dH},  we find that it
flows to the trivial ground state, such as $\mathbb{T}^{111}_{111}=1$. This implies that none of the phases in the phase diagram Fig. \ref{phase diagram} has the intrinsic topological order.

Then, we apply the SP-QSRG to the 2D AKLT model \eq{2dH}.  We find the ground state by using iTEBD method with bond dimension $\chi=2$  to study the Hamiltonian \eq{2dH}. After performing the spectral decomposition for the tensor $\mathbb{T}$ in Fig. \ref{o-2DRG-1}, we see the degeneracy pattern of eigenvalues (with $1$, $3$, $5$, $7$ etc fold degeneracy). Therefore, the wavefunction has $SO(3)$ symmetry and could possibly be a weak SPT phase with $SO(3)$ and translational symmetry.

Since the original physical index per site is in the half-integer representation, i.e., spin $3/2$, so we should require the physical indices of the tensors on each site in the intermediate steps to be also in the half-integer representation to ensure the SPT. The simplest way is to interpret the physical index of $\tilde{T}$ as comprised of two spin $3/2$. Thus, we can understand the degeneracy pattern of the singular values as follows: $4\otimes 4 = 1 \oplus 3 \oplus 5 \oplus 7$. Accordingly, we should enlarge the Hilbert space of the physical index of the tensor $\tilde{T}$ and split it by the rule of the measured quantum number in such a decomposition.

%The simplest way is to interpret the physical index of $\tilde{T}$ as comprised of four spin $1/2$ ``partons". Thus, we can understand the degeneracy pattern of the singular values as follows: $2\otimes 2 \otimes 2 \otimes  2= 1 \oplus 1 \oplus 3 \oplus 3 \oplus 3 \oplus 5$. Accordingly, we should enlarge the Hilbert space of the physical index of the tensor $\tilde{T}$ and split it by the rule of the measured quantum number in such a decomposition, e.g., for the 2D AKLT exact state, fill empty elements for two of the three $3$-representations and one of the two $1$-representations.

\begin{figure}[ht]
\center{\epsfig{figure=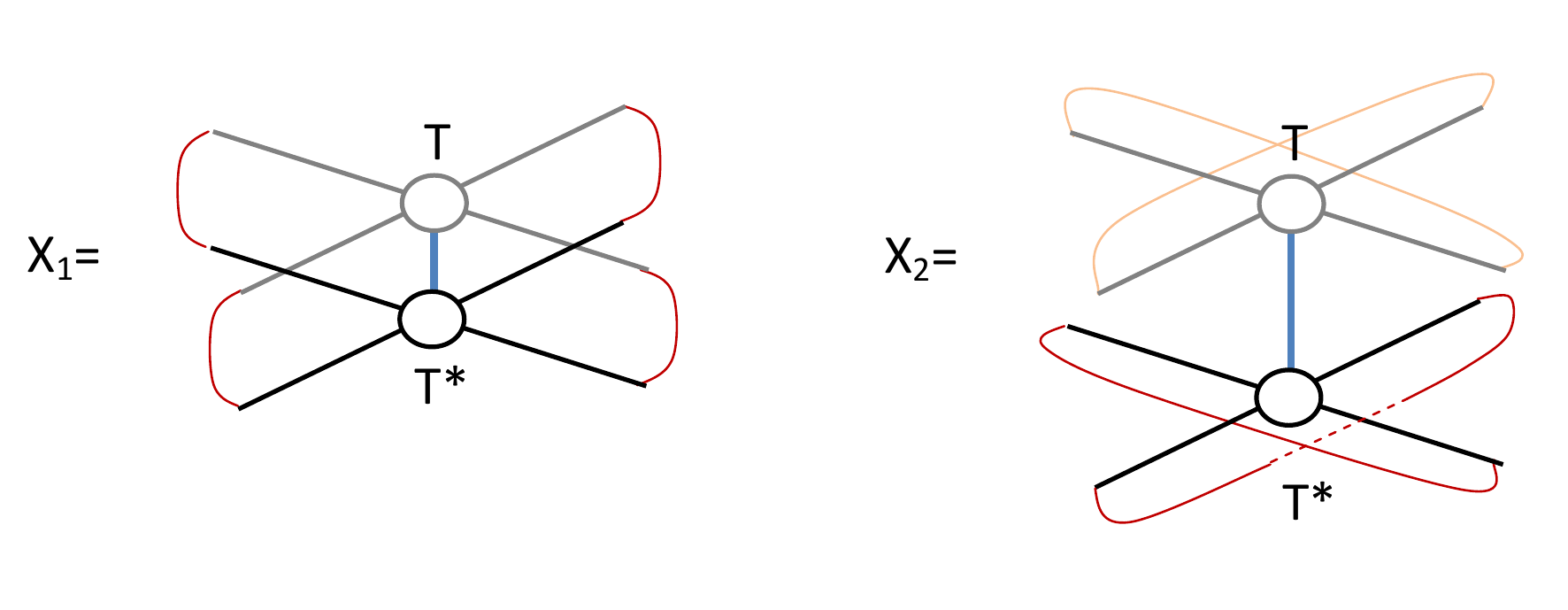,angle=0,width=7cm}}
\caption{The quantity $X_2/X_1$ obtained by taking the ratio of the contraction value of the double tensor in two different ways. $X_2/X_1$ is invariant under changing scale and gauge transformation. It can be used to distinguish different fixed-point tensors.  }
   \label{x2x1}
\end{figure}

In order to see the invariant structure of the fixed point up to change of scale and basis, we calculate some invariant quantity. We define a quantity that is given by the ratio of $X_2$ and $X_1$ (see Fig. \ref{x2x1}), as follows:
 \begin{align}\label{X2}
&X_1=\sum_{s,\alpha,\beta,\gamma,\delta} T^s_{\alpha,\beta,\gamma,\delta} T^{*s}_{\alpha,\beta,\gamma,\delta} \\ \notag
&X_2=\sum_{s,\alpha,\beta,\gamma,\delta} T^s_{\alpha,\beta,\alpha,\beta}  T^{*s}_{\gamma,\delta,\gamma,\delta}
\end{align}
where $s$ is the physical index and $\alpha,\beta,\gamma,\delta$ are bond indices. It is simple to verity that this quantity is invariant under the change in scale and the basis transformation. In the Fig \ref{x2x1}, we show the tensor representation on square lattice. We can merge two neighbor sites on honeycomb lattice to form a new tensor representation on square lattice. It is easy to calculate this quantity on honeycomb lattice. For 2D AKLT state (exact TPS), we have $X_2/X_1=0.28$.

\begin{figure}[ht]
\center{\epsfig{figure=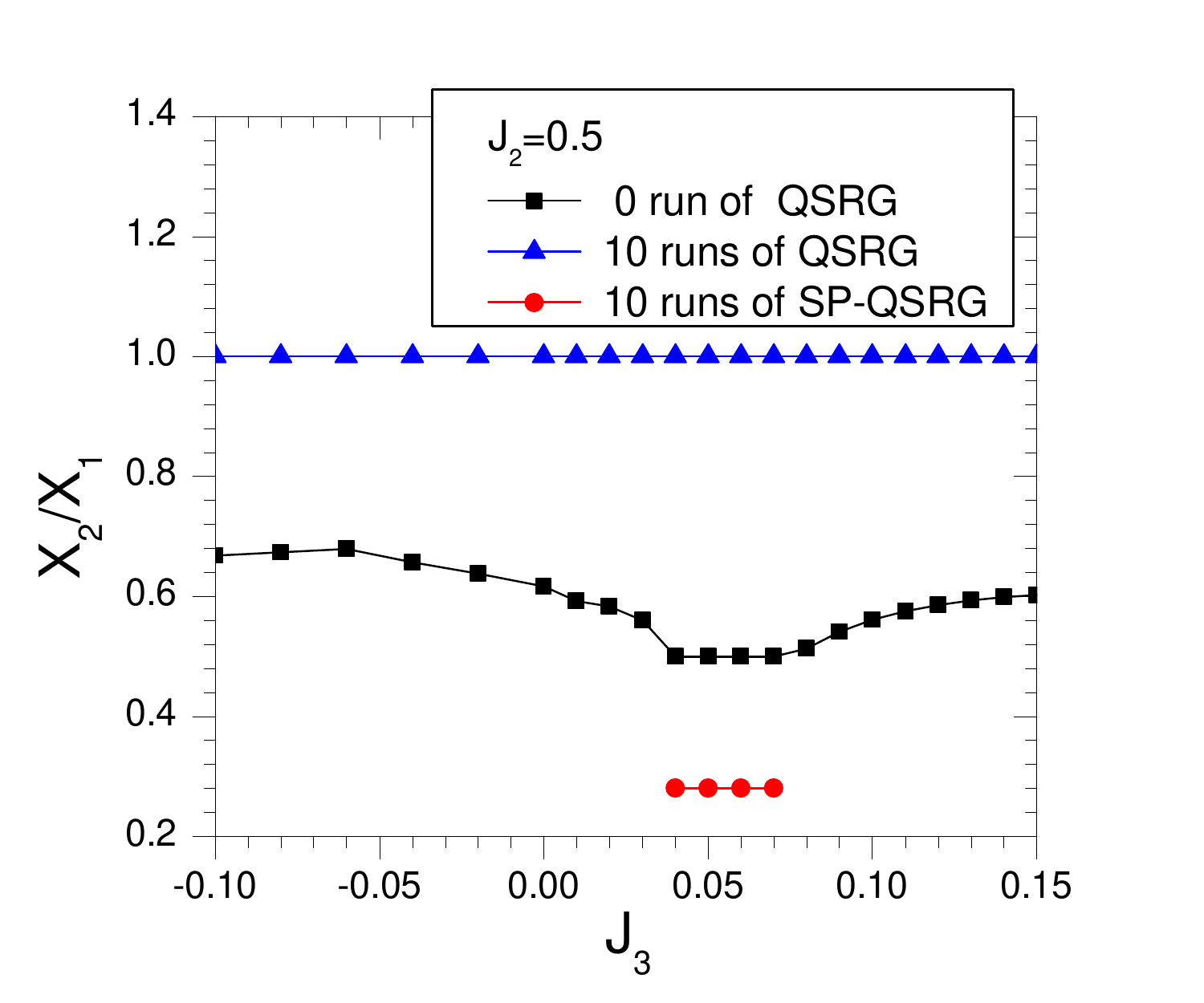,angle=0,width=7cm}}
\caption{The quantity $X_2/X_1$  for tensors under the renormalization flow by tuning $J_3$ with $J_1$ and $J_2=0.5$ with (circle) and without(triangle) preserving symmetry. The square is the $X_2/X_1$ of ground state without doing QSRG.}
   \label{x2x1_aklt}
\end{figure}

We find the ground state by using iTEBD method with bond dimension $\chi=2$ to study the Hamiltonian (\ref{2dH}) and  calculate the quantity $X_2/X_1$  in the process of QSRG. The results are shown in Fig. \ref{x2x1_aklt}  by tuning $J_3$ with $J_1=1$ and $J_2=0.5$ fixed, from which we can see that all points in the nonsymmetry breaking region belong to the same phase. This phase diagram can match the  Fig. \ref{J20.5} with order parameters. Before doing QSRG, the $X_2/X_1$  as functions of $J_3$ seems to be a smooth function (see squares in Fig. \ref{x2x1_aklt}). If we don't preserve symmetry, all of state will flow to trivial state, $X_2/X_1=1$ (see triangles in Fig. \ref{x2x1_aklt}). Then, preserving $SO(3)$ symmetry, for the AKLT phase region, $X_2/X_1$ tends to 0.28 (see circles in Fig. \ref{x2x1_aklt}), and  they flow to the same fixed point. This is true not only for parameters drawn in Fig. \ref{x2x1_aklt}, but for the whole AKLT phase region as well. In other regions, symmetry is broken and we cannot apply the SP-QSRG algorithm and hence no data points are shown.

Let us stress again that the quantity $X_2/X_1$ of the fixed point tensor is a unique value. It is an invariant quantity under the change in scale and the basis transformation of the indices. As a result, it is a good parameter to distinguish different fixed point tensors. Other invariant quantities for fixed point tensors can be constructed using similar ideas.

In order to show that this is the SPT fixed point, we apply the SP-QSRG procedure to the 2D dimer state, represented by the $T_a$ and $T_b$ sublattice tensors
 \begin{align}\label{X2}
& T_{a,000}^{000}=1,\qquad T_{a,111}^{111}=-1  \\ \notag
& T_{a,011}^{011}=1,\qquad T_{a,110}^{110}=1,\qquad T_{a,101}^{101}=1   \\ \notag
& T_{a,001}^{001}=-1,\qquad T_{a,010}^{010}=-1,\qquad T_{a,100}^{100}=-1.
\end{align}

 \begin{align}\label{X2_b}
& T_{b,000}^{111}=1,\qquad T_{b,111}^{000}=1  \\ \notag
& T_{b,011}^{100}=1,\qquad T_{b,110}^{001}=1,\qquad T_{b,101}^{010}=1   \\ \notag
& T_{b,001}^{110}=1,\qquad T_{b,010}^{101}=1,\qquad T_{b,100}^{011}=1.
\end{align}

This state is the valence bond solid with each pair of ``partons" on the neighboring sites forming singlet state. In the language of TPS, the projection operator is the identity. We find that it flows to the same fixed point with the same $X_2/X_1$ value and. This then demonstrates the power of our SP-QSRG and also justifies the SPT of the AKLT phase.

%%%%%%%%%%%%%%%%%%%%%%%%%%%%%%%%%%%%%%%%%%%%%%%%%%%%%%%%%%%%%%%%%%%%%%%%%%%%%%%%%%%%%%%%%%%%%%%%%%%%%%%%%%%%%%%%%%%%%%%%%%%%%%%%%%%%%%%%%%%%%%%%
%%%%%%%%%%%%%%%%%%%%%%%%%%%%%%%%%%%%%%%%%%%%%%%%%%%%%%%%%%%%%%%%%%%%%%%%%%%%%%%%%%%%%%%%%%%%%%%%%%%%%%%%%%%%%%%%%%%%%%%%%%%%%%%%%%%%%%%%%%%%%%%%

\section{Conclusion}
\label{conclusion}

The experimental search of exotic topological phases has attracted great attention. To motivate and assist the search, it is important to predict theoretically possible models which could have interesting topological properties.  As most of the exactly solvable higher dimensional models with (symmetry protected) topologically ordered ground states involve non-realistic interactions, numerical simulation is usually necessary to find the possible topologically ordered phases in more realistic models.

In this paper we have aimed at such a goal in devising a numerical algorithm based on the matrix and tensor product state representation of the ground states. Our algorithm of Symmetry Protected Quantum State Renormalization Group transformation can flow a ground state wave function to its fixed-point form in the same SPT phase. The fixed-point form of the MPS and TPS is usually simple and universal due to the removal of irrelevant short range entanglements so that it can be used to identify the SPT. This algorithm is the modified version of the original quantum state renormalization group algorithm\cite{1drg,2drg}, which enforces the symmetry constraints protecting the SPT orders. The key to the success of the algorithm is to make sure that the symmetry representation of degrees of freedom per site remains in the same class of projective representation during the RG process.

We have considered the 1D and 2D AKLT phases as examples in testing our SP-QSRG algorithm. These models have on-site $SO(3)$ symmetry so that only integer spin degrees of freedom can be added or removed when doing the RG transformation. Our numerical implementation of the algorithm on these states yield satisfying results by finding that they all flow the same fixed-point states as the dimer states do.  This confirms that the AKLT and the dimer states are in the same SPT class for the 1D and 2D cases. Moreover, we are able to see a clear phase transition between trivial and nontrivial SPT phases as they flow to different fixed point tensors.

We note that the splitting procedure in the RG algorithm involves certain arbitrariness and is not unique. Different procedure corresponds to different ways of removing local entanglement and leads to different forms of fixed point tensor. It is desirable to fix this ambiguity and find an efficient way to determine the  splitting procedure which removes local entanglment in an optimal way.

Finally, future directions involve exploring similar algorithm for the identification of SPT order with spatial symmetry (e.g. reflection in 1D), `strong' SPT order (without translation symmetry protection) in two and higher dimensions, and symmetry enriched topological order. Also it would be interesting to apply these algorithms to realistic models in search of such exotic topological phenomena.

{\bf Note added:} After finishing the first version of our paper, we learned of the work ``Symmetry Protected Entanglement Renormalization" done by Singh and Vidal \cite{SPMERA} considering the same issue as ours but in the context of MERA.

\acknowledgments
 We would like to acknowledge discussion with Xiao-Gang Wen and Frank Pollmann. XC is supported by the Miller Institute for Basic Research in Science at UC Berkeley.  FLL is supported by Taiwan's NSC grants (grant NO. 100-2811-M-003-011 and 100-2918-I-003-008) and he also thanks the support of NCTS.

%%%
%\section*{Acknowledgements}
%%%%%%%%%%%%%%%%

\appendix
\section{Projective Representation}\label{app a}
\label{prorep}

Matrices $u(g)$ form a projective representation of symmetry group $G$ if
\begin{align}
 u(g_1)u(g_2)=\om(g_1,g_2)u(g_1g_2),\ \ \ \ \
g_1,g_2\in G.
\end{align}
Here $\om(g_1,g_2)$'s are $U(1)$ phase factors, which is called the
factor system of the projective representation. The factor system satisfies
\begin{align}
 \om(g_2,g_3)\om(g_1,g_2g_3)&=
 \om(g_1,g_2)\om(g_1g_2,g_3),
\end{align}
for all $g_1,g_2,g_3\in G$.
If $\om(g_1,g_2)=1$, this reduces to the usual linear representation of $G$.

A different choice of pre-factor for the representation matrices
$u'(g)= \bt(g) u(g)$ will lead to a different factor system
$\om'(g_1,g_2)$:
\begin{align}
\label{omom}
 \om'(g_1,g_2) =
\frac{\bt(g_1g_2)}{\bt(g_1)\bt(g_2)}
 \om(g_1,g_2).
\end{align}
We regard $u'(g)$ and $u(g)$ that differ only by a pre-factor as equivalent
projective representations and the corresponding factor systems $\om'(g_1,g_2)$
and $\om(g_1,g_2)$ as belonging to the same class $\om$.

Suppose that we have one projective representation $u_1(g)$ with factor system
$\om_1(g_1,g_2)$ of class $\om_1$ and another $u_2(g)$ with factor system
$\om_2(g_1,g_2)$ of class $\om_2$, obviously $u_1(g)\otimes u_2(g)$ is a
projective presentation with factor group $\om_1(g_1,g_2)\om_2(g_1,g_2)$. The
corresponding class $\om$ can be written as a sum $\om_1+\om_2$. Under such an
addition rule, the equivalence classes of factor systems form an Abelian group,
which is called the second cohomology group of $G$ and denoted as
$\cH^2[G,U(1)]$.  The identity element $1 \in \cH^2[G,U(1)]$ is the class that
corresponds to the linear representation of the group.

\section{Numerical results for solving 2-dimensional AKLT-like model}\label{app b}

\begin{figure}[ht]
\center{\epsfig{figure=
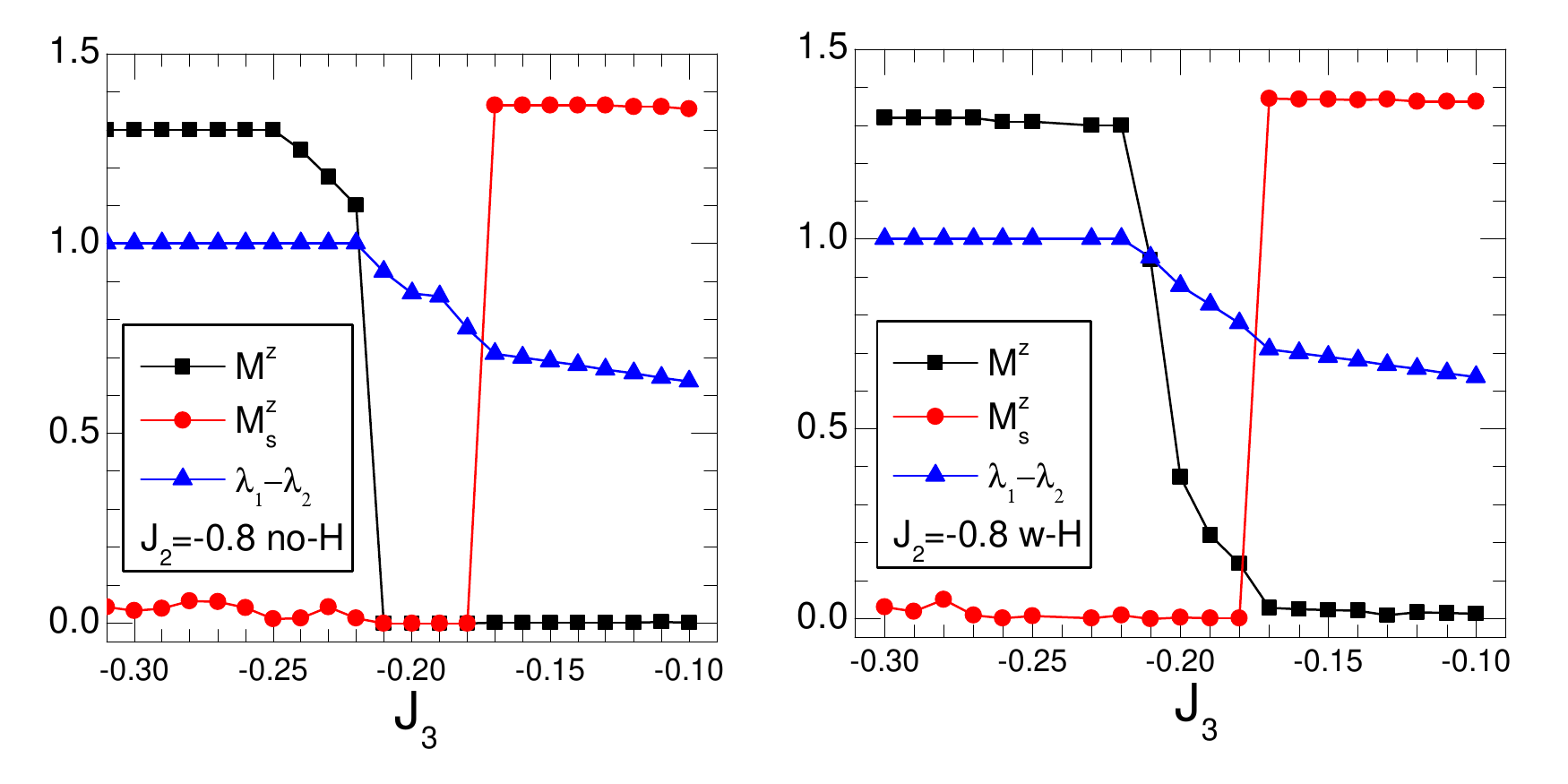,angle=0,width=9cm}}
\caption{A typical phase
diagram for the magnetization $M^z$ and the staggered magnetization
$M^z_s$ by tuning $J_3$ with $J_1=1$ and $J_2=-0.8$ fixed. We fix
the bond dimension $\chi=4$ and $D_{cut}=24$ in this numerical
calculation.The left one is without annealing but the right one is
with annealing by tuning a small magnetic field in the z-direction
when evaluating the ground state.  This annealing kill the sudden
jump of $M^z$. We also calculate the  energy per site for
$J_3=-0.2$. The energies with  and without the annealing are
$-7.303$ and $-7.150$, respectively. }
   \label{J2=-0.8-compare}
\end{figure}

We adopt the method of simple update \cite{itebd} to solve the ground state of \eq{2dH} numerically. Then, we apply the method of TRG \cite{trg} to evaluate the relevant order parameters and delineate the phase diagrams as shown in Fig. \ref{J20.5} and \ref{J2=-0.8-compare}.  The total phase diagram of this model as shown in Fig. \ref{phase diagram}.

\begin{figure}[ht]
\center{\epsfig{figure=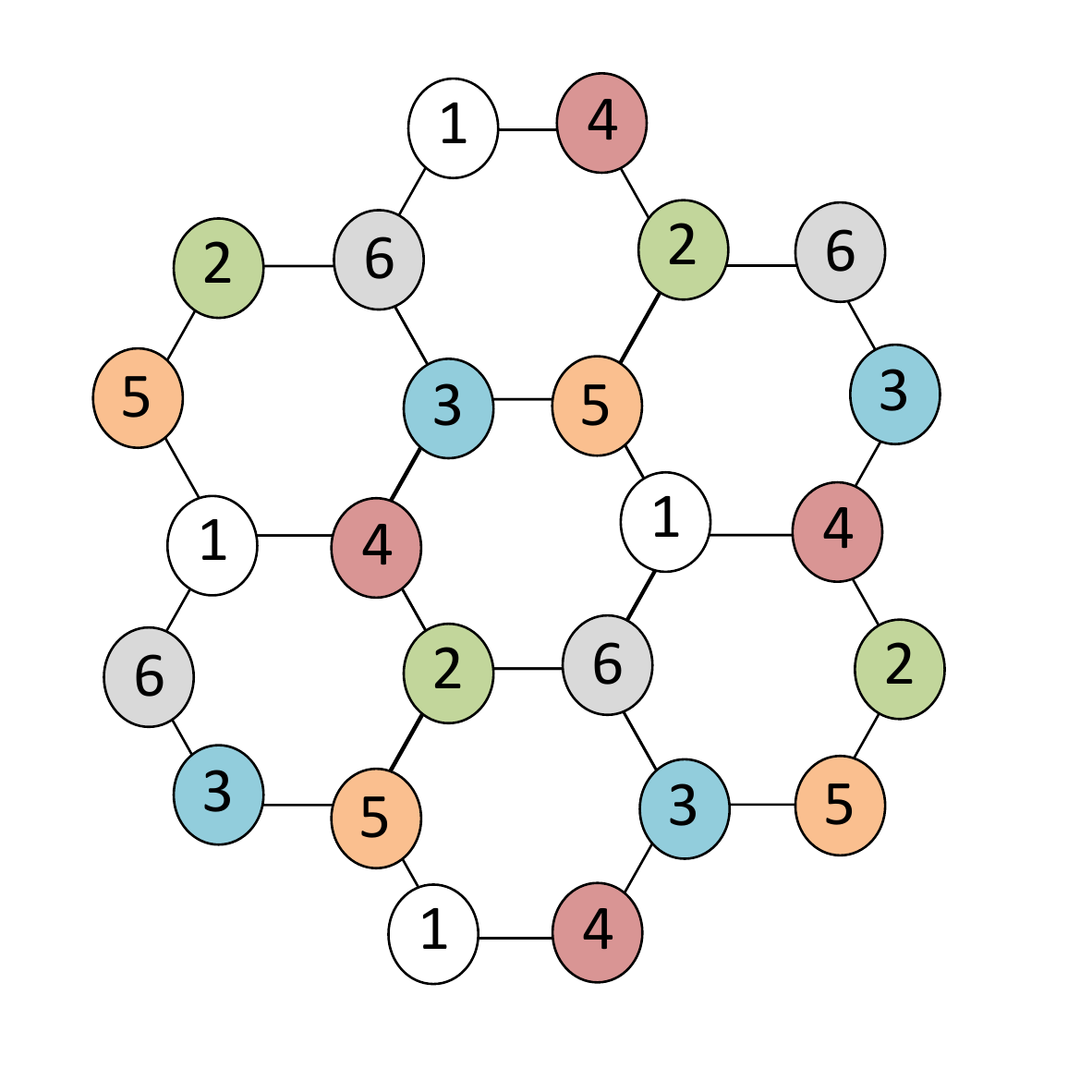,angle=0,width=5cm}}
\caption{Diagrammatic representation of the TPS on the honeycomb
lattice. The tensors $T^{(i)}$, with $i=1,2,..,6$ on the site
labeled by $i$ has three bond indices and one physical index.}
   \label{tensor-con}
\end{figure}

We assume the translational invariance for the TPS ansatz for the ground state of \eq{2dH}. Using the simple update method we can solve the TPS numerically. One should caution that the unit cell used in the simple update is a honeycomb, which is different from the acyclic tree of coordination number equal to $3$. So, when performing each step of the simple update, we need to
update the six kinds of tensors and nine bonds as shown in Fig. \ref{tensor-con}.

\begin{figure}[ht]
\center{\epsfig{figure=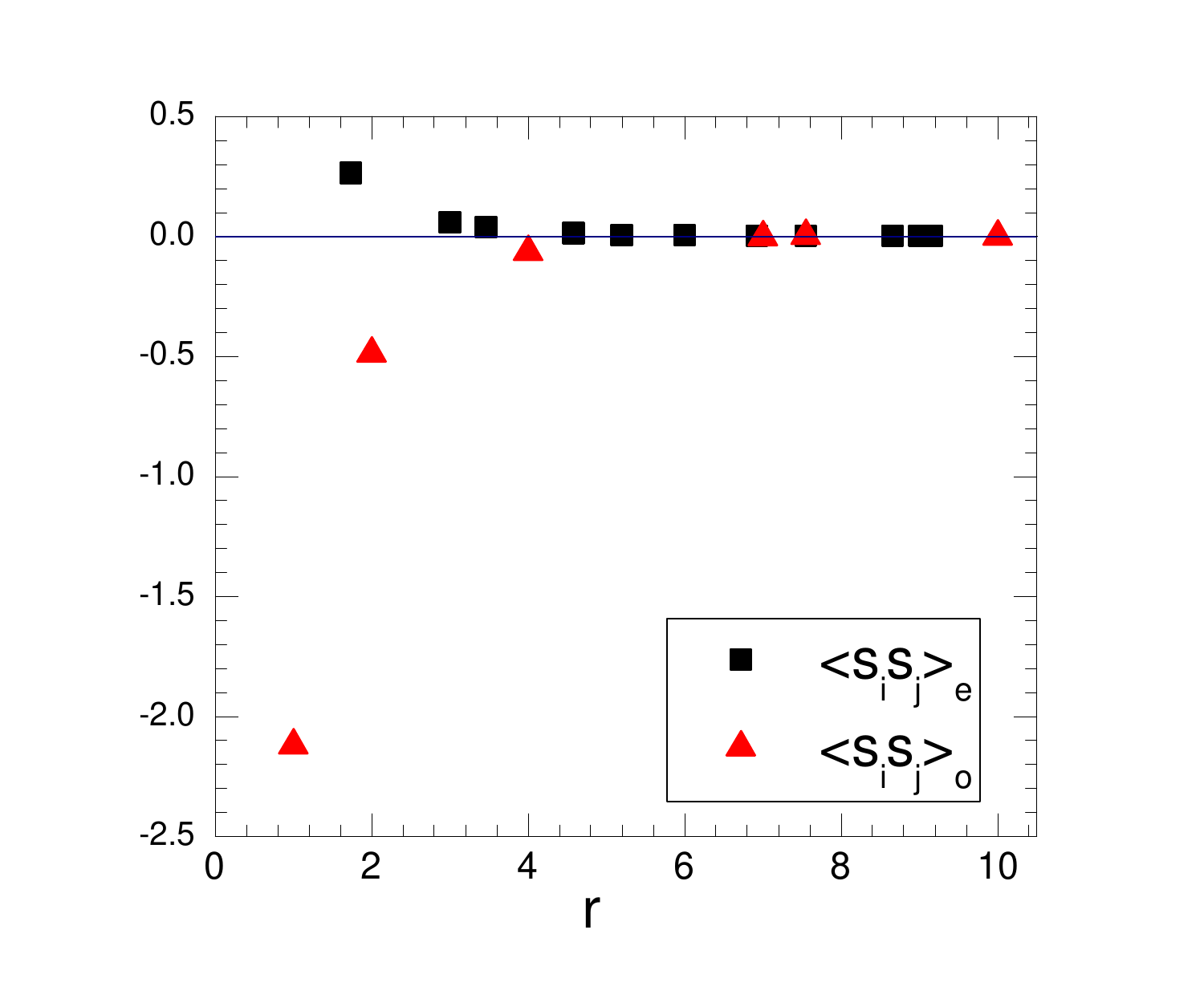,angle=0,width=8cm}}
\caption{The spin-spin correlation function as a function of
$r=|i-j|$. In this numerical calculation, $\chi=4$ and $D_{cut}=24$ are used.}
   \label{correlation}
\end{figure}

Based on the numerical solution from simple update we can further use TRG method to evaluate the expectation value of the magnetization denoted by $M^z$ and the staggered magnetization $M^z_s$. They are the order parameters for the ferromagnetic and the Neel phases, respectively. In our numerical calculation, we consider the bond dimension up to $\chi=5$ and keep $D_{cut}\geq \chi^2$ which is a cutoff of the merging bond dimension to  ensure the accuracy of the TRG calculation.

A typical  numerical result for $M^z$ and $M^z_s$ are shown in Fig. \ref{J20.5} with $J_1=1$ and $J_2=0.5$, which shows a quantum phase transition around $J_3=0$ from the N$\acute{e}$el phase to the AKLT phase as we decrease $J_3$. Note that the difference between largest two singular values also indicates the quantum phase transition. If we further decrease $J_3$, we will reach the ferromagnetic phase.  We also plot another numerical results with $J_2=-0.8$ in Fig. \ref{J2=-0.8-compare}. It shows that the N$\acute{e}$el order will suddenly drop to zero and remain zero after that. On the other hand, the  $M^z$ is zero but suddenly start to grow.  However, if anneal the result by tuning small magnetic field, then it shows that  $M^z$ grows gradually. This indicates there are only the ferromagnetic and the N$\acute{e}$el phases in this regime.

  We  also calculate the spin-spin correlation function at the AKLT point
in the Fig \ref{correlation}. It shows the exponentially-decay behavior as expected for a gapped system like the AKLT model.

\section{Fixed point tensor for 2-dimensional AKLT phae}\label{app c}

 In this Appendix, we write down the fixed point tensor of states in 2D AKLT phase with bond dimension $\chi=2$.

 In general, the explicit form of this tensor will depend upon the choice of the bases. Fortunately,
 the 2D AKLT phase and dimer phase have closed-form tensor representation with  $\chi=2$. By this way of performing SP-QSRG on the AKLT state,
we finally arrive the following fixed-point double tensor:
\begin{align}\notag
 & \mathbb{T}_{111,111}=0.05;\\ \notag
 & \mathbb{T}_{112,112}=0.044;\qquad  \mathbb{T}_{121,121}=0.044;\qquad  \mathbb{T}_{211,211}=0.044;\\ \notag
 & \mathbb{T}_{221,221}=0.036;\qquad  \mathbb{T}_{212,212}=0.036;\qquad  \mathbb{T}_{122,122}=0.036;\\
  & \mathbb{T}_{222,222}=0.05.
     \label{2daklt-fp}
\end{align}

The reasons for unique fixed point tensor are exactly similar in the 1D and 2D cases. We checked that this is also the fixed point double tensor of the 2D
dimer state.

Besides, we also apply the SP-QSRG to the other numerical ground states with $\chi=2$ in the
AKLT phases,such as $ J_2=116/243$,$ J_3=16./243+0.01$ and  $ J_2=0.8$,$ J_3=0.15$, they all flow to the same fixed point state given by
\eq{2daklt-fp}. This then demonstrates the power of our SP-QSRG and also justifies the SPT of the AKLT phase.

 \end{document}